\documentclass[showpacs,preprintnumbers,amsmath,amssymb,prb]{revtex4}

\def\XXint#1#2#3{{\setbox0=\hbox{$#1{#2#3}{\int}$}
     \vcenter{\hbox{$#2#3$}}\kern-.5\wd0}}

\newcommand{\cj}{{\cal J}}

\usepackage{inputenc}
\usepackage{graphicx}
\usepackage{subfigure}
\usepackage{bm}

\begin{document}

\title{Collective excitations in spin-$\frac12$ magnets through bond-operator formalism designed both for paramagnetic and ordered phases}

\author{A.\ V.\ Syromyatnikov$^{1,2}$}
\email{asyromyatnikov@yandex.ru}
\affiliation{$^1$National Research Center "Kurchatov Institute" B.P.\ Konstantinov Petersburg Nuclear Physics Institute, Gatchina 188300, Russia}
\affiliation{$^2$St.\ Petersburg State University, 7/9 Universitetskaya nab., St.\ Petersburg, 199034
Russia}

\date{\today}

\begin{abstract}

We present a bond-operator theory (BOT) suitable for description both magnetically ordered phases and paramagnetic phases with singlet ground states in spin-$\frac12$ magnets. This technique allows to trace evolution of quasiparticles across the transition between the phases. Some elementary excitations described in the theory by separate bosons appear in conventional approaches as bound states of well-known quasiparticles (magnons or triplons). The proposed BOT provides a regular expansion of physical quantities in powers of $1/n$, where $n$ is the maximum number of bosons which can occupy a unit cell (physical results correspond to $n=1$). Two variants of BOT are suggested: for two and for four spins in the unit cell (two-spin and four-spin BOTs, respectively). We consider spin-$\frac12$ Heisenberg antiferromagnet (HAF) on simple square lattice bilayer by the two-spin BOT. The ground-state energy $\cal E$, the staggered magnetization $M$, and quasiparticles spectra found within the first order in $1/n$ are in good {\it quantitative} agreement with previous results both in paramagnetic and in ordered phases not very close to the quantum critical point between the phases. By doubling the unit cell in two directions, we discuss spin-$\frac12$ HAF on square lattice using the suggested four-spin BOT. We identify the magnon and the amplitude (Higgs) modes among fifteen spin-2, spin-1, and spin-0 quasiparticles arisen in the theory. Magnon spectrum, $\cal E$, and $M$ found in the first order in $1/n$ are in good {\it quantitative} agreement with previous numerical and experimental results. We observe a special moderately damped spin-0 quasiparticle ("singlon" for short) whose energy is smaller than the energy of the Higgs mode in the most part of the Brillouin zone. By considering HAF with Ising-type anisotropy, we find that both Higgs and "singlon" modes stem from two-magnon bound states which merge with two-magnon continuum not far from the isotropic limit. We demonstrate that "singlons" appear explicitly in "scalar" correlators one of which describes the Raman intensity in $B_{1g}$ symmetry. The latter is expressed in the leading order in $1/n$ via the "singlon" Green's function at zero momentum which shows an asymmetric peak. The position of this peak given by the "singlon" energy coincides with the position of the so-called "two-magnon" peak observed experimentally in, e.g., layered cuprates. 

\end{abstract}


\maketitle

\section{Introduction}

Search and characterization of elementary excitations (quasiparticles) is of fundamental importance for the modern theory of strongly interacting many-body systems. A wealth of collective phenomena are discussed in terms of appropriate quasiparticles, interaction between them, and their decay into other elementary excitations. Then, the role is important of convenient and powerful theoretical approaches allowing to introduce and to operate with suitable elementary excitations. Theories are of particular importance relying on expansions around exactly solvable limits because they allow to describe accurately a certain area of parameter space. Examples include $1/N$-expansions, where $N$ is the number of flavors or the number of order-parameter components, $\epsilon$-expansions, where $\epsilon=d_c-d$, $d$ is the space dimension, and $d_c$ is the upper or lower critical dimension, and $1/S$-expansion, where $S$ is the spin value. \cite{auer,zinn,sachdev} Such theories provide in some cases even quantitatively accurate results far beyond the formal domain of their applicability. 

One of such approaches is $1/S$-expansion which is based on Holstein-Primakoff (or on Dyson-Maleev) spin transformation. \cite{auer,hp} It allows to describe magnetic systems in ordered phases in terms of elementary excitations named magnons. In most cases, one can find only a few first terms in $1/S$-series for observable quantities. It is well known, however, that even truncated $1/S$-series can provide surprisingly accurate results even when the formal condition of the theory applicability, $S\gg1$, is far from being fulfilled. The notable example is spin-$\frac12$ Heisenberg antiferromagnet (HAF) on square lattice. \cite{monous} $1/S$-expansion failed to work well near phase transitions when the nature of elementary excitations changes: e.g., near classical phase transitions or near quantum phase transitions (QPTs) when extra critical modes appear. 

The prominent example of the latter situation is QPT from magnetically ordered phase to dimerized phase with singlet ground state, when the amplitude (Higgs) mode comes into play. \cite{sachdev,chub,joshi2} The Higgs mode is one of fundamental collective excitations arisen in various systems with spontaneously broken continuous symmetry and corresponding to fluctuations of the order parameter amplitude (along with Goldstone excitations corresponding to fluctuations of the order parameter phase). \cite{higgs1} It is not convenient to take it into account within $1/S$-expansion because the amplitude mode arises in this technique as a pole of a two-magnon vertex. \cite{chub,joshi2} To obtain this pole one has to take into account infinite number of diagrams. The amplitude mode has attracted much attention recently as it bears close correspondence with Higgs modes in particle physics. \cite{higgs1} Deep in the ordered phase, the amplitude mode is a high-energy excitation with finite lifetime caused by decay into two Goldstone quasiparticles. Due to its damping, it is undetectable deep in the ordered phase in measurements of order-parameter correlators \cite{vis1,vis2,podol} (the longitudinal spin susceptibility in magnetic systems) while it is visible in scalar correlators \cite{podol} (many-spin, or bond-bond, correlators in magnetic systems \cite{podol,vojta,weidinger2015}). An advance in neutron experimental technique allows to observe it recently in $\rm TlCuCl_3$ near the pressure-induced QPT, where the Higgs mode is sharp. \cite{tlcucl1,tlcucl2} It has been proposed also that interaction between the amplitude mode and magnons is responsible for the roton-like minimum in magnon spectrum at ${\bf k}=(\pi,0)$ in spin-$\frac12$ HAF on square lattice. \cite{pow1,pow2} This minimum is not described quantitatively by standard analytical approaches including $1/S$-expansion (see Refs.~\cite{pow1,pow2,piazza,spinon,syromyat} and references therein). Then, it has been argued recently that an excitation by light of two Higgs quasiparticles is responsible for a shoulder-like anomaly in Raman intensity in $B_{1g}$ geometry arisen in some layered cuprates near the so-called "two-magnon" peak. \cite{weidinger2015} 

The amplitude mode has been discussed so far analytically either using field-theoretical approaches \cite{sachdev,podol,weidinger2015,vis1,vis2} or using bond-operator theories (BOTs) \cite{chub,joshi1,sommer}. Originally, some variants of bond-operator spin representations have been proposed to describe paramagnetic phases with singlet ground state. \cite{chubm1,bhatt,chub0,chub,joshi1,sommer,bond} BOTs have been also developed which are able to describe both the ordered and the dimerized phase (and QPT between the two). \cite{chub,joshi1,joshi2} There is a separate Bose-operator in such BOTs describing the Higgs excitation that makes these techniques much more precise and convenient compared with, e.g., $1/S$-expansion. \cite{chub} It is explicitly seen in these theories that the Higgs mode turns into a spin-0 excitation (one of triplet excitations called triplons) upon transition to the dimerized phase. A weakness of the majority of BOTs suggested so far is the absence of an expansion parameter (see Refs.~\cite{joshi1,joshi2} for an extended discussion). It has been overcome in dimerized phase in Ref.~\cite{chub} by introducing a formal parameter $n$ of maximum number of bosons which can occupy a bond (the ordered phase can be also considered by this technique near the quantum critical point (QCP) between the two phases in terms of "condensation" of triplons). A variant of BOT is proposed in Refs.~\cite{joshi1,joshi2} which allows to find observable quantities in both phases as series in powers of $1/d$. Results obtained by this approach in the first order in $1/d$ are in quantitative agreement with other numerical and analytical findings (see also below). \cite{vojta} A drawback of this technique is that it does not allow to calculate the Higgs mode damping.

BOT proved to be very useful in discussions of other elementary excitations which are normally treated as bound states of conventional quasiparticles. Thus, in our previous paper \cite{inem}, we discuss a QPT from fully polarized to a nematic phase in frustrated spin-$\frac12$ quasi-one-dimensional ferromagnet in strong magnetic field. The nematic phase appears in this system as a result of "condensation" of two-magnon bound states upon field decreasing. \cite{chubnem} We double the unit cell along the chain in Ref.~ \cite{inem} and develop a BOT which takes into account all spin degrees of freedom in each unit cell. Three bosonic quasiparticles arise in that technique two of which carry spin 1 and describe two parts of the conventional magnon mode. We argue in Ref.~ \cite{inem} that the third boson carries spin 2 and describes the two-magnon bound states which "condense" at QCP. The problem is exactly solvable within that formalism in the saturated phase. The presence of the bosonic mode in the theory which softens at QCP makes substantially standard the QPT consideration. \cite{inem}

One of the aims of the present paper is to show that there exist quasiparticles inside ordered phases whose role has not been fully clarified yet. We propose below BOTs for two and for four spins in the unit cell. The suggested spin representations are parametrized in such a way that they are suitable for consideration both ordered and paramagnetic phases. Thus, these approaches allow to trace evolution of elementary excitations across QPTs and on moving between different exactly solvable limits. These representations depend on the formal parameter $n$, the maximum number of bosons in the unit cell, in such a way that the theory allows a regular expansion in powers of $1/n$ (which differs, however, from the variant of $1/n$-expansion suggested in Ref.~\cite{chub} for the dimerized phase and the neighborhood of QCP). Remarkably, the spin commutation algebra is reproduced at any $n\ge1$ that guarantees, in particular, existence of Goldstone excitations in ordered phases with spontaneously broken continuous symmetry in any oder in $1/n$. Thus, we overcome the problem of many previous BOTs (see Refs.~\cite{joshi1,joshi2} for an extended discussion).

Indeed, the value of expansion in powers of $1/n$ might seem questionable in the physically relevant case of $n=1$ (as the value of $1/S$-expansion at $S\sim1$, though). Thus, after introduction of the spin representation for two spins in the unit cell in Sec.~\ref{spinrep2}, we discuss in Sec.~\ref{hafbi} in detail spin-$\frac12$ HAF on square lattice bilayer which has been well studied before by various methods. The latter circumstance provides a good opportunity to test the ability of the proposed formalism. We demonstrate in Sec.~\ref{hafbi} that the ground-state energy $\cal E$, staggered magnetization $M$, and quasiparticle spectra found within the first order in $1/n$ (and taken at $n=1$) are in good quantitative agreement with previous results not very close to QCP. Thus, the situation with the proposed $1/n$-expansion is very similar to that with $1/S$-expansion in spin-$\frac12$ HAF on square lattice, where corrections of the first order in $1/S$ give the main renormalization of observables. \cite{monous}

We introduce in Sec.~\ref{method4} the spin representation for four spins in the unit cell assuming for definiteness that the unit cell has the form of a plaquette. The theory is quite cumbersome in this case as it contains fifteen Bose-operators. We apply this formalism to spin-$\frac12$ HAF on square lattice in Sec.~\ref{ssl} by doubling the unit cell in two directions. Results of our calculation in Sec.~\ref{static4} of $\cal E$ and $M$ in the first order in $1/n$ is in good and in excellent quantitative agreement with previous findings, respectively. 

We consider in Sec.~\ref{excit4} the evolution of quasiparticles spectra from the exactly solvable limit of isolated plaquettes to HAF on the square lattice. There is a QPT on this way from the paramagnetic to the ordered phase which helps us to identify the Higgs mode among other spin-0 excitations. We find that along with high-energy spin-2, spin-1, and spin-0 excitations there is a special spin-0 quasiparticle which is purely singlet in the disordered phase. Such singlet excitations appeared in previous BOTs with two spins in the unit cell as singlet bound states of two triplons. \cite{singlon1,singlon2} Singlet excitations in paramagnetic phases can be called singlons for short. We call their counterpart "singlons" (in quotes) in the ordered phase, where they are no more singlet. We introduce the Ising-type anisotropy to the system and consider the exactly solvable Ising limit to demonstrate that the amplitude mode and "singlons" stem from two-magnon bound states which enter into the two-magnon continuum not far from the isotropic limit. 

We calculate in Sec.~\ref{excithaf} quasiparticles spectra in spin-$\frac12$ HAF on square lattice within the first order in $1/n$ and demonstrate that both amplitude and "singlons" modes are moderately damped. We find that "singlons" lie below the amplitude mode in the major part of the Brillouin zone (BZ). Magnon spectrum is in good quantitative agreement with previous numerical and experimental results even around ${\bf k}=(\pi,0)$.

We demonstrate in Sec.~\ref{ramansec} that "singlons" are not visible in dynamical spin structure factors but they appear explicitly in "scalar" correlators one of which describes the Raman intensity in $B_{1g}$ symmetry. We show that the latter is expressed in the leading order in $1/n$ via the "singlon" Green's function at zero momentum which possesses an asymmetric peak at $\omega\approx2.74J$, where $J$ is the exchange coupling constant. The peak position (but not the width) coincides with the position of the "two-magnon" peak in Raman intensity observed experimentally in, e.g., layered cuprates. The spectral weight of this peak is comparable with that of the "two-magnon" peak obtained before at $\omega\approx3.3J$ within $1/S$-expansion in the ladder approximation. However, an analysis is needed in further orders in $1/n$ to describe the experimental data in every detail which we carry out elsewhere.

In the forthcoming paper, \cite{weunp} we will discuss using the proposed formalism spin-$\frac12$ $J_1$--$J_2$ HAF on square lattice, where the frustrating $J_2>0$ exchange coupling is added between next-nearest-neighbor spins. We will demonstrate, in particular, that the "singlon" spectrum moves down and the "singlon" damping decreases upon $J_2$ increasing (the spectrum remains gapped at all $J_2$, however). "Singlons" become long-lived quasiparticles and their spectrum nearly merges with the magnon spectrum in the most part of BZ at $J_2\approx0.3J_1$. Singlons are purely singlet low-energy excitations in the paramagnetic phase (i.e., at  $0.4J_1<J_2<0.6J_1$).

We provide a summary and a conclusion in Sec.~\ref{conc}. One appendix is added with details of the analysis.

\section{Bond-operator formalism for two spins in the unit cell}
\label{method2}

\begin{figure}
\includegraphics[scale=0.8]{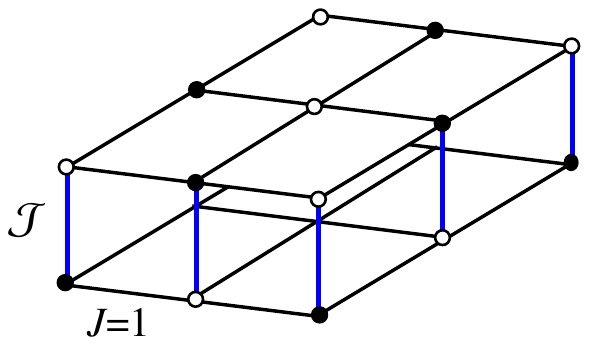}
\caption{Spin-$\frac12$ HAF on square lattice bilayer considered using the suggested bond-operator formalism with two spins in the unit cell. Lattice sites belonging to two different sublattices in the ordered phase are distinguished by color (spins ${\bf S}_1$ and ${\bf S}_2$ in representation \eqref{trans}). 
\label{system}}
\end{figure}

We develop in this section BOT for two spins 1/2 in the unit cell bearing in mind for definiteness the spin-$\frac12$ HAF on simple square lattice bilayer shown in Fig.~\ref{system} whose Hamiltonian has the form
\begin{equation}
\label{hambi}
{\cal H} = \cj\sum_j	{\bf S}_{1j}{\bf S}_{2j} 
+ 
\sum_{ \langle i,j \rangle }\left( {\bf S}_{1i}{\bf S}_{1j} + {\bf S}_{2i}{\bf S}_{2j} \right),
\end{equation}
where $\cj\ge0$, indexes 1 and 2 enumerate layers (spins in the unit cell), the intralayer exchange coupling constant is set to be equal to unity, and $\langle i,j \rangle$ denote nearest-neighbor sites in a layer. This model has been well studied before by various methods (see, e.g., Refs.~\cite{vojta,joshi1,joshi2} and references therein) that provides a good opportunity to test the ability of the proposed formalism. It is well known, in particular, that the QPT arises in this model at $\cj=\cj_c\approx2.52$ from the N\'eel ordered state to the dimerized phase. 

\subsection{Spin representation}
\label{spinrep2}

To derive a representation for spins ${\bf S}_{1j}$ and ${\bf S}_{2j}$ in the $j$-th unit cell, we introduce three Bose-operators $a_j^\dagger$, $b_j^\dagger$, and $c_j^\dagger$ which create three mutually orthogonal spin states from a vacuum $|0\rangle$ as follows: 
\begin{eqnarray}
\label{states}
&&|0\rangle = \cos\alpha\left|\uparrow\downarrow\rangle\right.
- \sin\alpha\left|\downarrow\uparrow\rangle\right.,\nonumber\\
a^\dagger |0\rangle &=& |a\rangle = \left|\uparrow\uparrow\rangle\right.,\\
b^\dagger |0\rangle &=& |b\rangle = \left|\downarrow\downarrow\rangle\right.,\nonumber\\
c^\dagger |0\rangle &=& |c\rangle = \sin\alpha\left|\uparrow\downarrow\rangle\right.
+ \cos\alpha\left|\downarrow\uparrow\rangle\right.,\nonumber
\end{eqnarray}
where $\alpha$ is a real parameter. It is seen that the vacuum $|0\rangle$ is a singlet state at $\alpha=\pi/4$ whereas the N\'eel order
(i.e., $\langle0|S_1^z|0\rangle = -\langle0|S_2^z|0\rangle \ne0$)
arises when $\sin\alpha\ne\cos\alpha$. Parameter $\alpha$ allows to connect smoothly the singlet and the N\'eel ordered phases. We propose the following representation for ${\bf S}_{1j}$, ${\bf S}_{2j}$, and $({\bf S}_{1j}{\bf S}_{2j})$:
\begin{subequations}
\label{trans}
\begin{eqnarray}
\label{s1+}
S_{1j}^+ &=& -a_j^\dagger P_j \sin\alpha 
+ P_j b_j \cos\alpha 
+ c_j^\dagger b_j \sin\alpha 
+ a_j^\dagger c_j \cos\alpha, \\
S_{1j}^- &=& -P_j a_j \sin\alpha 
+ b_j^\dagger P_j \cos\alpha 
+ b_j^\dagger c_j \sin\alpha 
+ c_j^\dagger a_j \cos\alpha, \\
\label{s1z}
S_{1j}^z &=& n\frac{\cos2\alpha}{2} 
+\frac{\sin 2\alpha}{2} \left(P_j c_j + c_j^\dagger P_j\right) 
+ a_j^\dagger a_j\sin^2\alpha 
- b_j^\dagger b_j\cos^2\alpha 
- c_j^\dagger c_j \cos2\alpha,\\
S_{2j}^+ &=& a_j^\dagger P_j \cos\alpha 
- P_j b_j \sin\alpha 
+ c_j^\dagger b_j \cos\alpha 
+ a_j^\dagger c_j \sin\alpha, \\
S_{2j}^- &=& P_j a_j \cos\alpha 
- b_j^\dagger P_j \sin\alpha 
+ b_j^\dagger c_j \cos\alpha 
+ c_j^\dagger a_j \sin\alpha, \\
\label{s2z}
S_{2j}^z &=& -n\frac{\cos2\alpha}{2} 
-\frac{\sin 2\alpha}{2} \left(P_j c_j + c_j^\dagger P_j\right) 
+ a_j^\dagger a_j\cos^2\alpha 
- b_j^\dagger b_j\sin^2\alpha 
+ c_j^\dagger c_j \cos2\alpha, \\
\label{ss}
\left({\bf S}_{1j}{\bf S}_{2j}\right) &=& - n^2\frac{1+2\sin2\alpha}{4} 
+ n\frac{\cos 2\alpha}{2} \left(P_j c_j + c_j^\dagger P_j\right) 
+ n\frac{1+\sin2\alpha}{2}\left(a_j^\dagger a_j + b_j^\dagger b_j\right) 
+ nc_j^\dagger c_j \sin2\alpha,
\end{eqnarray}
\end{subequations}
where
\begin{equation}
\label{proj}
P_j = \sqrt{n-a_j^\dagger a_j-b_j^\dagger b_j-c_j^\dagger c_j}
\end{equation}
is a projector on the physical subspace (consisting of states with no more than $n$ bosons in a unit cell) and $n=1$. It is easy to verify that operators in left-hand sides of Eqs.~\eqref{trans} act on spin states defined in Eqs.~\eqref{states} as operators in right-hand sides if $n=1$ (see also Table~\ref{table1}). An algorithm can be easily formulated to construct Eqs.~\eqref{trans} from the result of action of spin operators on states \eqref{states}. This algorithm (which can be programmed, e.g., in Mathematica Software) can be easily generalized to the case of more than two spins in the unit cell (see below). 

\begin{table}
\caption{Results of action of spin operators ${\bf S}_{1j}$, ${\bf S}_{2j}$, and $({\bf S}_{1j}{\bf S}_{2j})$ on spin states defined in Eqs.~\eqref{states}.
\label{table1}
}
\begin{ruledtabular}
\begin{tabular}{|c|c|c|c|c|c|}
  { }      & $S^z_1$ & $S^z_2$ & $S^+_1$ & $S^+_2$ & $({\bf S}_1{\bf S}_2)$ \\
\hline
$|0\rangle$ & $\displaystyle \frac{\cos2\alpha}{2}|0\rangle + \frac{\sin2\alpha}{2}|c\rangle$ 
& $\displaystyle -\frac{\cos2\alpha}{2}|0\rangle - \frac{\sin2\alpha}{2}|c\rangle$ 
& $-\sin\alpha|a\rangle$
& $\cos\alpha|a\rangle$
& $\displaystyle - \frac{1 + 2\sin2\alpha}{4} |0\rangle + \frac{\cos 2\alpha}{2}|c\rangle$  \\
$|a\rangle$ & $\displaystyle \frac12|a\rangle$ 
& $\displaystyle \frac12|a\rangle$ 
& $0$
& $0$
& $\displaystyle \frac14|a\rangle$ { } \\
$|b\rangle$ & $\displaystyle -\frac12|b\rangle$ 
& $\displaystyle -\frac12|b\rangle$  
& $\displaystyle \cos\alpha|0\rangle+\sin\alpha|c\rangle$ 
& $\displaystyle -\sin\alpha|0\rangle+\cos\alpha|c\rangle$ 
& \qquad $\displaystyle \frac14|b\rangle$  \\
$|c\rangle$ & $\displaystyle \frac{\sin2\alpha}{2}|0\rangle - \frac{\cos2\alpha}{2}|c\rangle $ 
& $\displaystyle - \frac{\sin2\alpha}{2}|0\rangle + \frac{\cos2\alpha}{2}|c\rangle $ 
& $\cos\alpha|a\rangle$
& $\sin\alpha|a\rangle$
& $\displaystyle \frac{\cos 2\alpha}{2}|0\rangle - \frac{1 - 2\sin2\alpha}{4} |c\rangle $ 
\end{tabular}
\end{ruledtabular}
\end{table}

It can be verified straightforwardly that for any $\alpha$ and $n\ge1$ representation \eqref{trans} reproduces the spin commutation algebra of operators ${\bf S}_{1j}$ and ${\bf S}_{2j}$ (i.e., $[S^\delta_{1j},S^\beta_{1j}]=i\epsilon_{\delta\beta\gamma}S^\gamma_{1j}$, $[S^\delta_{2j},S^\beta_{2j}]=i\epsilon_{\delta\beta\gamma}S^\gamma_{2j}$, and $[S^\delta_{1j},S^\beta_{2j}]=0$) and $({\bf S}_{1j}{\bf S}_{2j})$ given by Eq.~\eqref{ss} commutes with ${\bf S}_{1j}+{\bf S}_{2j}$. Notice that projector $P_j$ could contain $n-a_j^\dagger a_j-b_j^\dagger b_j-c_j^\dagger c_j$ in any positive power. It is for the spin algebra fulfillment that the power is equal to $1/2$ in Eq.~\eqref{proj}. Parameter $n$ can be considered arbitrary in all derivations with the Bose-analog of the spin Hamiltonian. However, only the case of $n=1$ has the physical meaning. It is seen that similar to the Holstein-Primakoff representation \cite{hp} Eqs.~\eqref{trans} have zero matrix elements between states from the Hilbert subspace with no more than $n$ bosons in the unit cell ("physical" subspace) and states with more than $n$ bosons ("unphysical" subspace).  Besides, it is shown below that the constant term in the Bose-analog of the spin Hamiltonian is of the order of $(1/n)^{-2}$, terms linear in Bose-operators are ${\cal O}((1/n)^{-3/2})$, bilinear terms are of $(1/n)^{-1}$ order, etc.
\footnote{Notice that we put factors $n^2$ and $n$ before the first and the last three terms in Eq.~\eqref{ss}, respectively, in order to make these terms of proper orders in $1/n$ in the Bose-analog of the spin Hamiltonian.} 
Then, expressions for physical observables can be obtained as series in the formal parameter $1/n$ and $n$ plays the role very much similar to the spin value $S$ in Holstein-Primakoff transformation.

Parameter $\alpha$ is to be found by minimization of the ground-state energy. In the singlet phase, $\alpha=\pi/4$ and Eqs.~\eqref{s1+}--\eqref{s2z} are equivalent to the spin representation suggested in Ref.~\cite{chub} for consideration of dimerized states. 
\footnote{Eqs.~\eqref{s1+}--\eqref{s2z} transform directly to Eqs.~(7) from Ref.~\cite{chub} by putting $\alpha=\pi/4$ and replacing $a$ by $-a$ and $c$ by $-c$ (a trivial unitary transformation of Bose-operators).}
Then, Eqs.~\eqref{s1+}--\eqref{s2z} is a generalization of that representation which is able to describe both singlet and magnetically ordered phases as well as transitions between them. However, our representation \eqref{ss} of operator $({\bf S}_{1j}{\bf S}_{2j})$ differs from that in Ref.~\cite{chub}, where $({\bf S}_{1j}{\bf S}_{2j})$ is expressed using Eqs.~\eqref{s1+}--\eqref{s2z} at $n=1$. As a result, in the singlet phase, the $1/n$-expansion suggested in the present paper differs from the variant of $1/n$-expansion proposed in Ref.~\cite{chub}. We find it more convenient to derive Bose-analogs of all spin operators in the unit cell (including $({\bf S}_{1j}{\bf S}_{2j})$) using the same procedure described above: it allows to make all terms in the Hamiltonian containing the same number of Bose-operators to be of the same order in $1/n$.

It should be stressed that the reproduction of the spin commutation algebra of operators ${\bf S}_{1j}$ and ${\bf S}_{2j}$ by Eqs.~\eqref{trans} and the commutativity of $({\bf S}_{1j}{\bf S}_{2j})$ (see Eq.~\eqref{ss}) with ${\bf S}_{1j}+{\bf S}_{2j}$ guarantee the existence of Goldstone excitations in the ordered phase within any order in $1/n$. 

\subsection{Spin-$\frac12$ HAF on square lattice bilayer}
\label{hafbi}

Substituting Eqs.~\eqref{trans} to Hamiltonian \eqref{hambi} and expanding the square root in projector \eqref{proj}, one obtains 
\begin{equation}
\label{hbose}
	{\cal H} = {\cal E} + \sum_{i=1}^\infty{\cal H}_i,
\end{equation}
where $\cal E$ is a constant and ${\cal H}_i$ stand for terms containing products of $i$ Bose-operators. In particular, we have
\begin{eqnarray}
\label{e0b}
\frac{{\cal E}}{N} &=& -\frac{n^2}{4} (2 + \cj + 2\cos 4\alpha + 2\cj \sin 2\alpha ), \\
\label{h1b}
\frac{{\cal H}_1}{\sqrt N} &=& n^{3/2} \frac{\cos 2\alpha}{2} (\cj - 4\sin 2\alpha ) \left(c_{\bf 0} + c_{\bf 0}^\dagger\right),\\
\label{h2}
{\cal H}_2 &=& \sum_{\bf k} \left(
A_{\bf k} \left( a_{\bf k}^\dagger a_{\bf k} + b_{\bf k}^\dagger b_{\bf k} \right) 
+ B_{\bf k} \left( a_{\bf k} b_{-\bf k} + a_{\bf k}^\dagger b_{-\bf k}^\dagger  \right)
+
E_{\bf k} c_{\bf k}^\dagger c_{\bf k} 
+ \frac{D_{\bf k}}{2} \left( c_{\bf k}^\dagger c_{-\bf k}^\dagger + c_{\bf k} c_{-\bf k} \right) 
\right),
\end{eqnarray}
where $N$ is the number of unit cells in the lattice and
\begin{eqnarray}
\label{coefb}
A_{\bf k} &=& 
\frac{n}{2} \left( \cj +\cj\sin 2\alpha + 4\cos^2 2\alpha - 2( \cos k_x + \cos k_y ) \sin 2\alpha \right),\nonumber\\
B_{\bf k} &=& n (\cos k_x + \cos k_y),\\
E_{\bf k} &=& 
n\left( \cj \sin 2\alpha + 4 \cos^2 2\alpha - (\cos k_x +\cos k_y )\sin^2 2\alpha \right),\nonumber\\
D_{\bf k} &=& -n (\cos k_x + \cos k_y) \sin^2 2\alpha.\nonumber
\end{eqnarray}

Minimization of ${\cal E}$ (see Eq.~\eqref{e0b}) gives the following value $\alpha_0$ of $\alpha$ in the leading order in $1/n$:
\begin{equation}
\label{a0b}
	\sin2\alpha_0 = \left\{
\begin{array}{ll}
\cj/4, & \mbox{ if } \cj < \cj_{c0}=4, \\
1, & \mbox{ if } \cj \ge \cj_{c0}.
\end{array}
\right.
\end{equation}
At $\alpha=\alpha_0$, linear term \eqref{h1b} vanishes in Hamiltonian and one obtains in the leading order in $1/n$ $\frac{\cal E}{2N} = -n^2(16+4\cj+\cj^2)/32$ and $-\frac38n^2 \cj$ for $\cj<\cj_{c0}$ and $\cj\ge\cj_{c0}$, correspondingly.
$M=\langle S_{1j}^z \rangle=-\langle S_{2j}^z \rangle=\frac n2\cos2\alpha_0$ 
is equal to $n \sqrt{16-\cj^2}/8$ and 0 when $\cj<\cj_{c0}$ and $\cj\ge\cj_{c0}$, respectively. Then, we find in agreement with previous results \cite{chub,joshi1,joshi2} that the system shows a QPT from the ordered to the dimerized phase at $\cj=\cj_c$, where $\cj_c=\cj_{c0}=4$ in the leading order in $1/n$. 

Bare spectra of $a$-, $b$-, and $c$-quasiparticles read as
\begin{eqnarray}
\epsilon_{0\bf k}^{(a)} &=& \epsilon_{0\bf k}^{(b)} = \sqrt{A_{\bf k}^2 - B_{\bf k}^2},\\
\epsilon_{0\bf k}^{(c)} &=& \sqrt{E_{\bf k}^2 - D_{\bf k}^2}.
\end{eqnarray}
In the ordered phase (i.e., at $\cj<\cj_{c0}$), $a$- and $b$- quasiparticles have a gapless spectrum and describe the conventional doubly degenerate magnon mode while $c$-quasiparticle represents the gapped amplitude (Higgs) mode. In the paramagnetic phase (i.e., at $\cj>\cj_{c0}$), all quasiparticles have the same gapped spectrum and represent the well-known triplons. 

\begin{figure}
\includegraphics[scale=0.4]{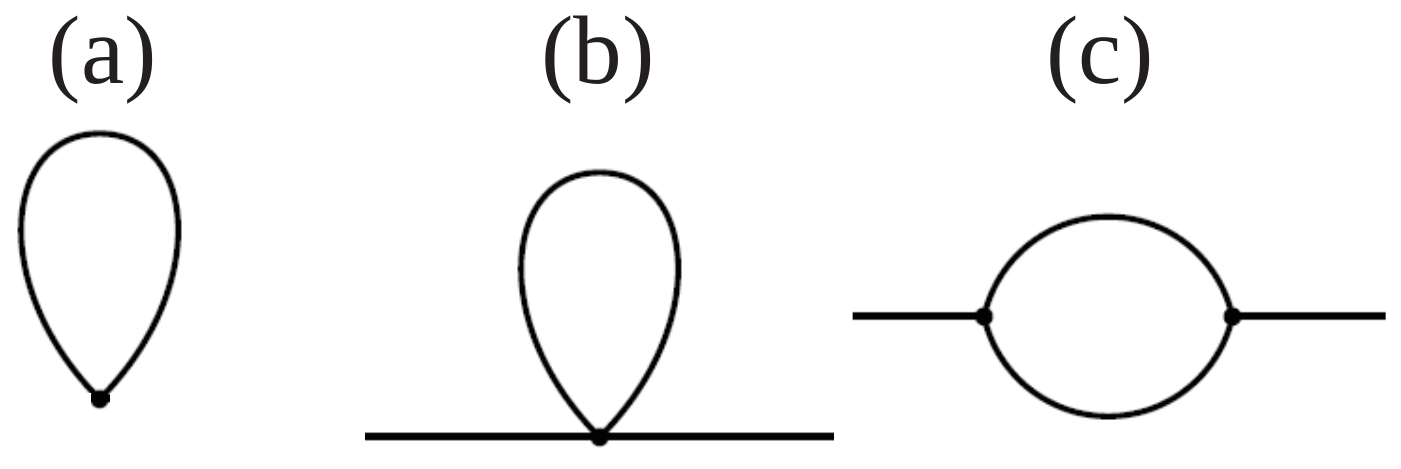}
\caption{Diagrams giving corrections of the first-order in $1/n$ to (a) the ground state energy and staggered magnetization, and (b), (c) to self-energy parts.
\label{diag}}
\end{figure}

First $1/n$ corrections to observable quantities can be found by the conventional diagrammatic technique. As in Refs.~\cite{inem,syromyat}, we use a technique which operates with anomalous Green's functions of the type 
$
G_{cc}(\omega,{\bf k}) = -i\int dt 
e^{i\omega t}	
\langle T c_{\bf k}(t)c_{-\bf k}(0)\rangle
$
and Green's functions of the "mixed" type
$
G_{ab}(\omega,{\bf k}) = -i\int dt 
e^{i\omega t}	
\langle T a_{\bf k}(t)b_{-\bf k}(0)\rangle
$ not involving Bogoliubov transformations. Then, one deals with sets of Dyson equations for the Green's functions within this approach. Such a technique is more compact and, thus, more convenient for cumbersome calculations. 

\begin{figure}
\includegraphics[scale=0.445]{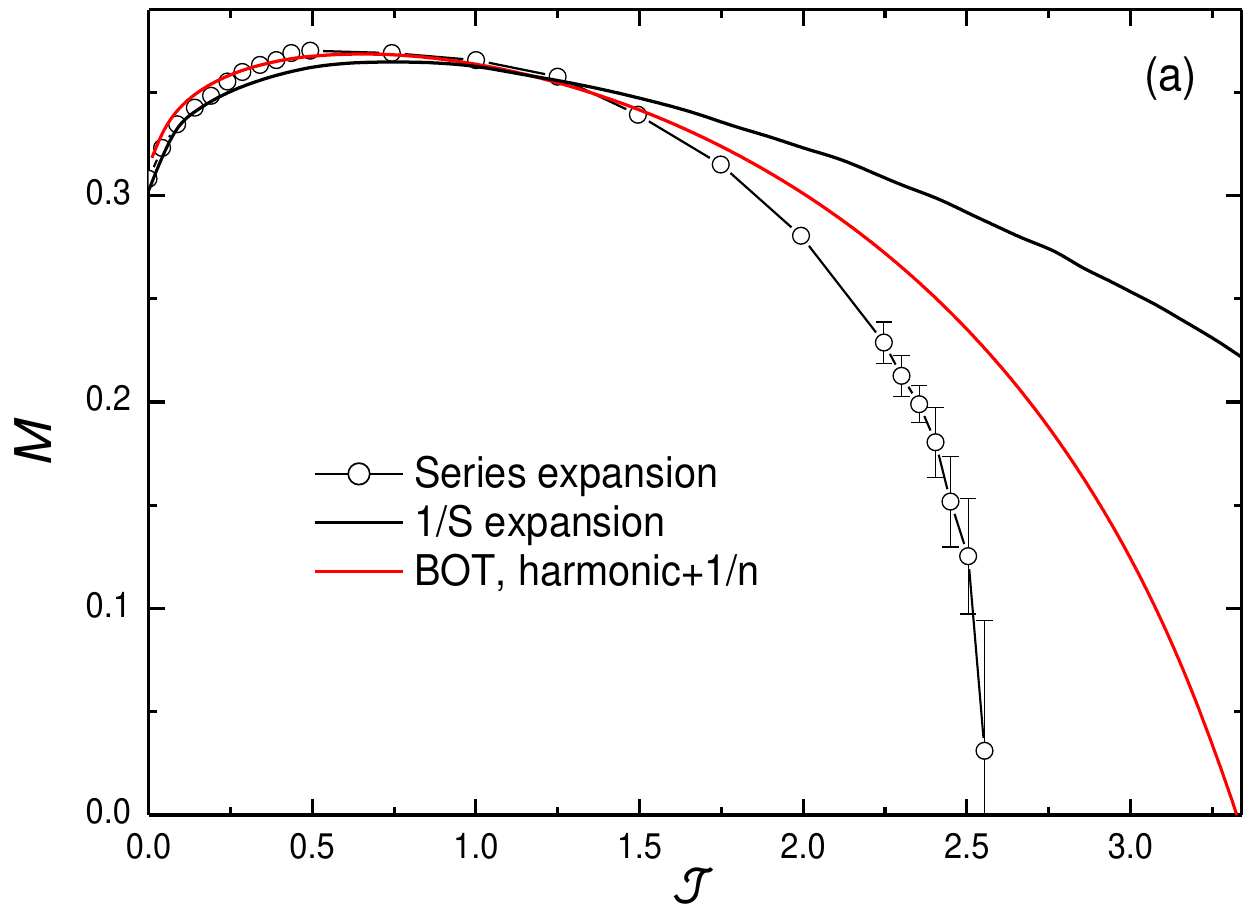}
\includegraphics[scale=0.41]{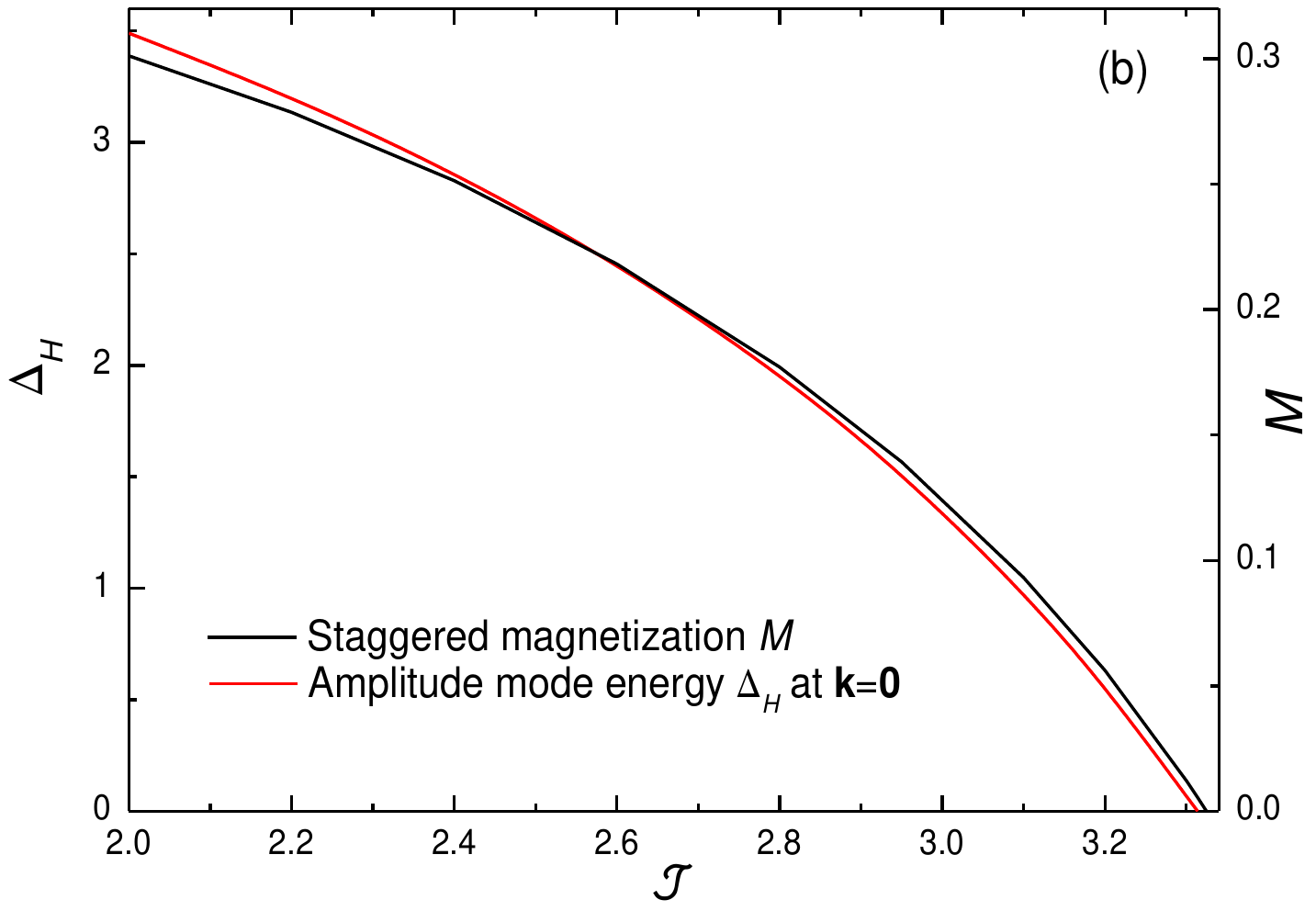}
\includegraphics[scale=0.41]{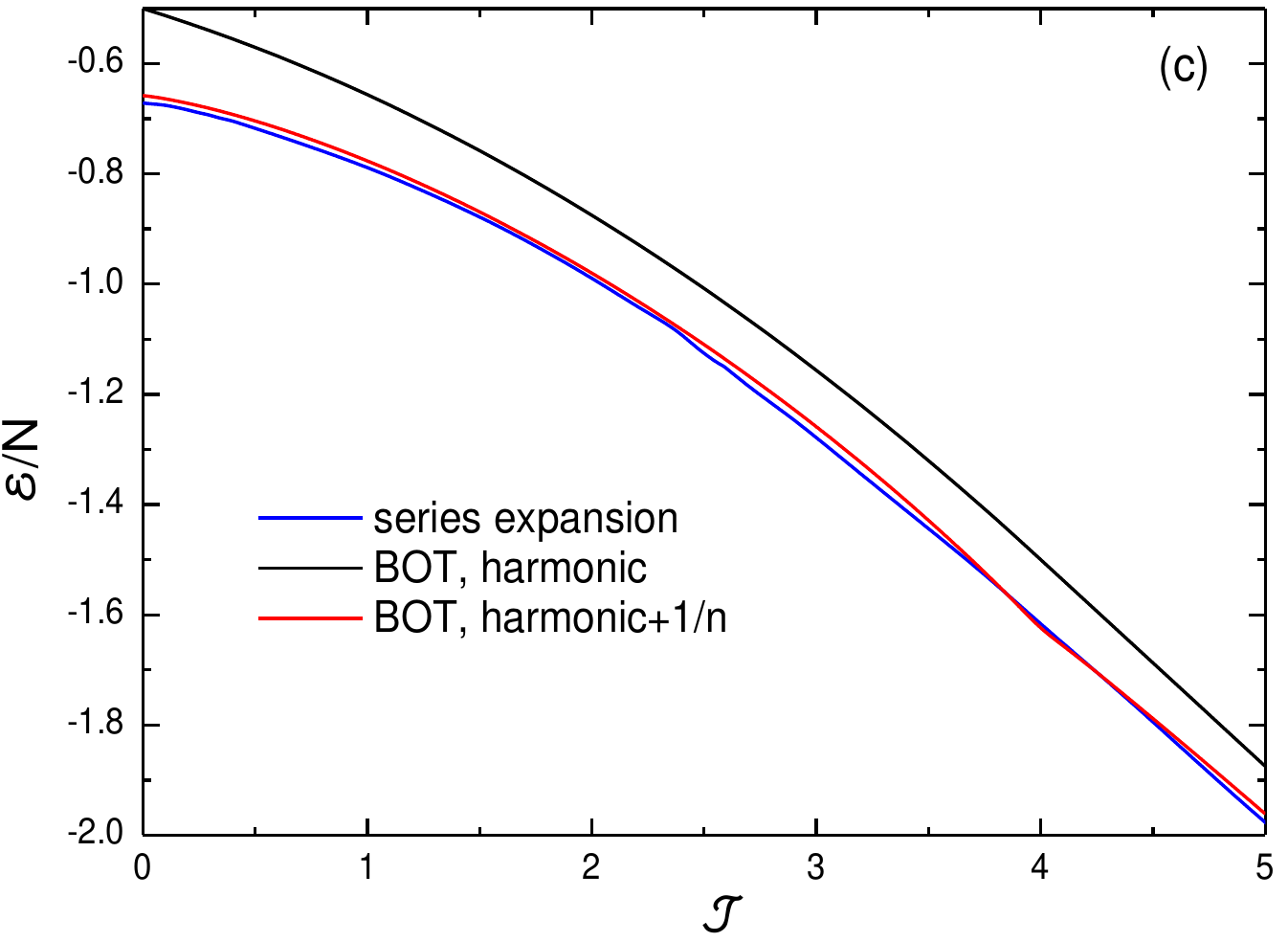}
\caption{(a) Staggered magnetization $M$ in spin-$\frac12$ HAF on square lattice bilayer found using the series expansion technique (taken from Ref.~\cite{weibi}), self-consistent spin-wave theory (taken from Ref.~\cite{chub}), and the bond-operator theory (BOT) in the first order in $1/n$ (present study). (b) $M$ and the gap in the spectrum of the amplitude (Higgs) mode $\Delta_H$ calculated within the first order in $1/n$ and taken at $n=1$. (c) The ground state energy $\cal E$ per spin obtained using the series expansion (taken from Fig.~3 in Ref.~\cite{weibi}) and the BOT.
\label{bimag}}
\end{figure}

${\cal H}_3$ and ${\cal H}_4$ terms in the Hamiltonian lead to diagrams of the first order in $1/n$ for self-energy parts shown in Figs.~\ref{diag}(b) and \ref{diag}(c). Besides, as soon as coefficients in the Hamiltonian depend on $\alpha$, renormalization of $\alpha$ contributes also to the renormalization of observables. By making all possible couplings of Bose operators in ${\cal H}_3$ (taken at $\alpha=\alpha_0$), one derives the first-order correction to ${\cal H}_1$ and obtains the correction to $\alpha_0$ from the requirement that ${\cal H}_1$ should vanish. Corrections to the ground state energy $\cal E$ and to the staggered magnetization $M=\langle S_{1j}^z \rangle$ come from the $\alpha$ renormalization and from all possible couplings of Bose operators in ${\cal H}_2$ and in bilinear terms in Eq.~\eqref{s1z}, respectively (see the diagram in Fig.~\ref{diag}(a)). For example, one obtains after simple calculations for $\cj=2$
\begin{eqnarray}
\alpha &=& \alpha_0 -0.1205 \frac1n,\\
\label{evalb}
\frac{\cal E}{2N} &=& -\frac78n^2 - 0.1049n,\\
\label{s1zvalb}
M &=& \langle S_{1j}^z \rangle=-\langle S_{2j}^z \rangle = \frac{\sqrt3}{4}n - 0.1318.
\end{eqnarray}
$M$ and the ground state energy per spin are presented in Figs.~\ref{bimag}(a) and \ref{bimag}(c) as functions of $\cj$ which have been found in the first order in $1/n$ and taken at $n=1$. As is seen, our results are consistent with those of series expansion technique \cite{weibi} and self-consistent spin-wave approach \cite{chub} not very close to the QCP $\cj_c\approx2.52$. We obtain that $M$ vanishes at 
\begin{equation}
\label{jcbi}
	\cj_{c1} = 4 - 0.6752\frac1n
\end{equation}
that gives $\cj_{c1}\approx3.3248$ at $n=1$ (nearly the same value of $\cj_{c1}\approx3.3684$ was obtained \cite{vojta,joshi1,joshi2} within the first order in $1/d$-expansion at $d=2$).

\begin{figure}
\includegraphics[scale=1.0]{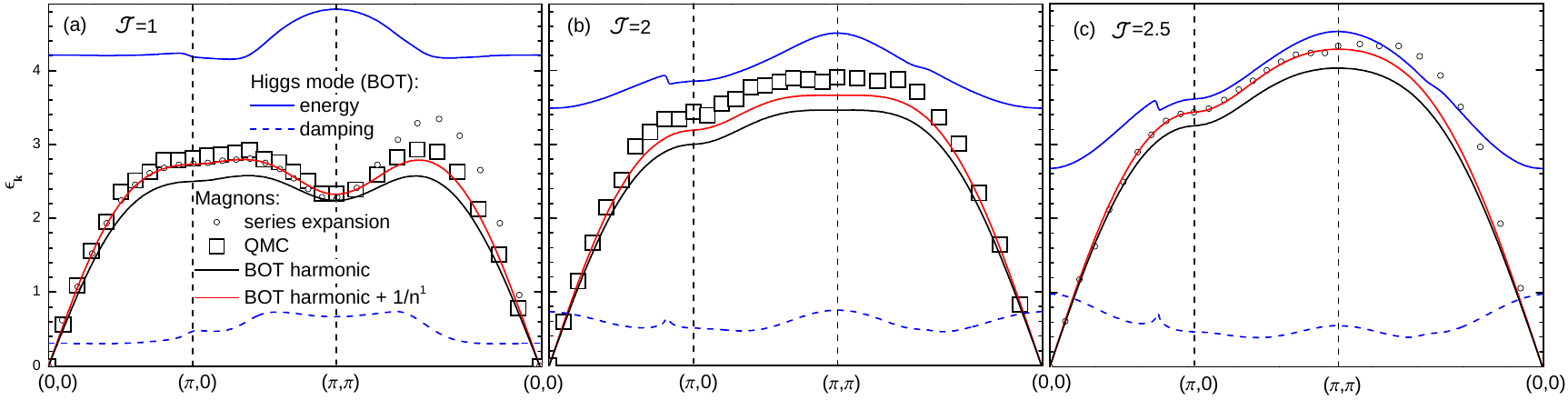}
\includegraphics[scale=1.03]{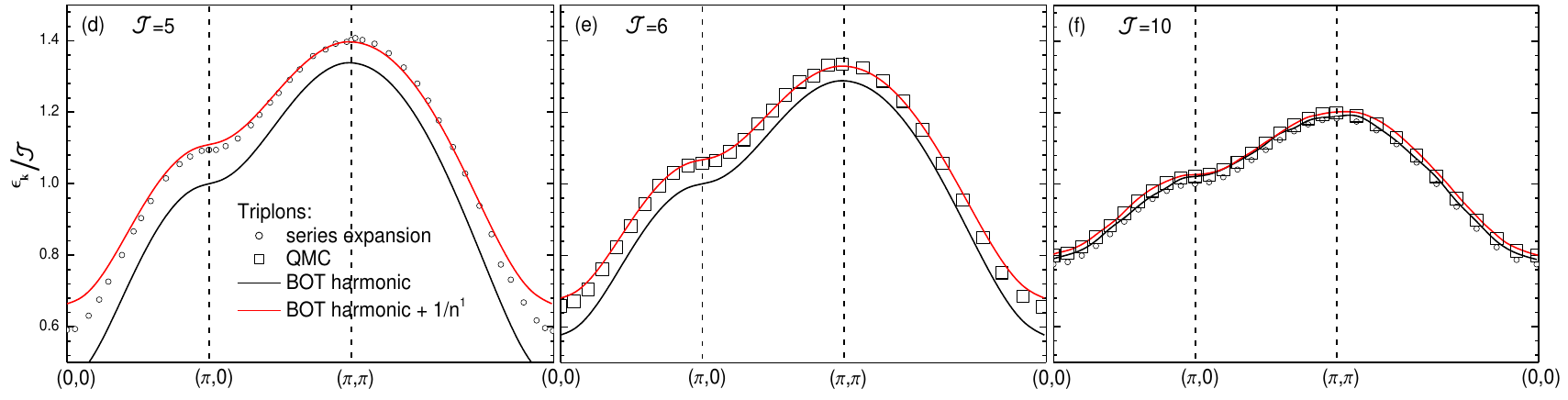}
\caption{(a)--(c) Spectra of elementary excitations (magnons and the amplitude (Higgs) mode) in spin-$\frac12$ HAF on square lattice bilayer in the ordered phase. Results are obtained using the series expansion technique (taken from Ref.~\cite{weibi}), Quantum Monte-Carlo (QMC) calculation on a sample with $2\times20\times20$ sites (taken from Ref.~\cite{vojta}), and the bond-operator theory (BOT) developed in the present paper. (d)--(f) Spectra of elementary excitations (triplons) in the dimerized state. The estimated uncertainty of quasiparticles energies in QMC data is indicated by the symbol size.
\label{specsqb}}
\end{figure}

Bare and renormalized spectra of quasiparticles are presented in Fig.~\ref{specsqb} for some $\cj$ values both in ordered and in disordered phases. It is seen that the magnon and triplon spectra found within the first order in $1/n$ are in good agreement with available previous numerical results obtained using QMC and series expansion not very close to the QCP ($\cj_c\approx2.52$). Notice that the magnon spectrum remains gapless in the ordered phase in the first order in $1/n$ as it must be. The amplitude mode acquires a damping due to the decay on two magnons described by the diagram shown in Fig.~\ref{diag}(c). Spikes in the Higgs mode damping accompanied by abrupt changes in its energy is the appearance of the Van Hove singularities from the two-particle density of states (similar anomalies were observed, e.g., in magnon spectra in the first order in $1/S$ in non-collinear magnets \cite{zhito,chub_triang}). The amplitude mode damping is overestimated near QCP in the first order in $1/n$ because bare spectra are used to calculate it. However its energy $\Delta_H$ at $\bf k=0$ vanishes nearly together with the order parameter (see Fig.~\ref{bimag}(b)). The slightly different values of $\cj$ at which $M$ and $\Delta_H$ vanishes is, evidently, a result of the restriction of $1/n$-expansion by the first terms. Magnon and triplon energies found in Ref.~\cite{vojta} in the first order in $1/d$ are also consistent with QMC data presented in Fig.~\ref{specsqb}.

\begin{figure}
\includegraphics[scale=0.9]{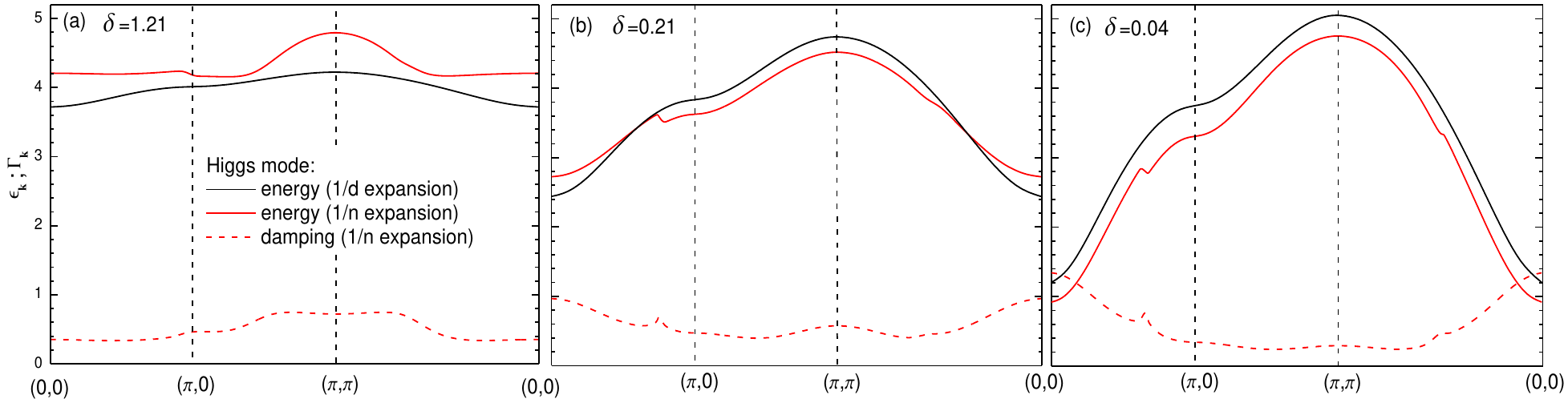}
\caption{Spectra of the amplitude (Higgs) mode in spin-$\frac12$ HAF on square lattice bilayer obtained in the first order in $1/d$ (Refs.~\cite{joshi2,vojta}) and in the first order in $1/n$ (present study). Results are shown for different values of parameter $\delta$ defined by Eq.~\eqref{delta} which measures the distance to the QCP in considered first orders in $1/n$ and $1/d$. 
\label{specHiggs}}
\end{figure}

A comparison is presented in Fig.~\ref{specHiggs} of the amplitude mode energy found within first orders of $1/d$- and $1/n$- expansions for three values of parameter 
\begin{equation}
\label{delta}
\delta = \frac{2}{\cj} - \frac{2}{\cj_{c1}}
\end{equation}
measuring the distance to the critical point within the considered order in $1/n$ or $1/d$. A good agreement is seen in Fig.~\ref{specHiggs} between the two analytical approaches. In turn, the results of the $1/d$-expansion presented in Fig.~\ref{specHiggs} are consistent with corresponding QMC data for the same value of $\delta$, as it is shown in Ref.~\cite{vojta} (see panels for $g=1$, 2, and 2.4 in Fig.~7 of Ref.~\cite{vojta}).

To conclude this section, we point out that first $1/n$-corrections give the main renormalization of observable quantities not very close to QCP. Consideration of further order corrections is out of the scope of the present paper.

\section{Bond-operator formalism for four spins in the unit cell}
\label{method4}

\begin{figure}
\includegraphics[scale=0.6]{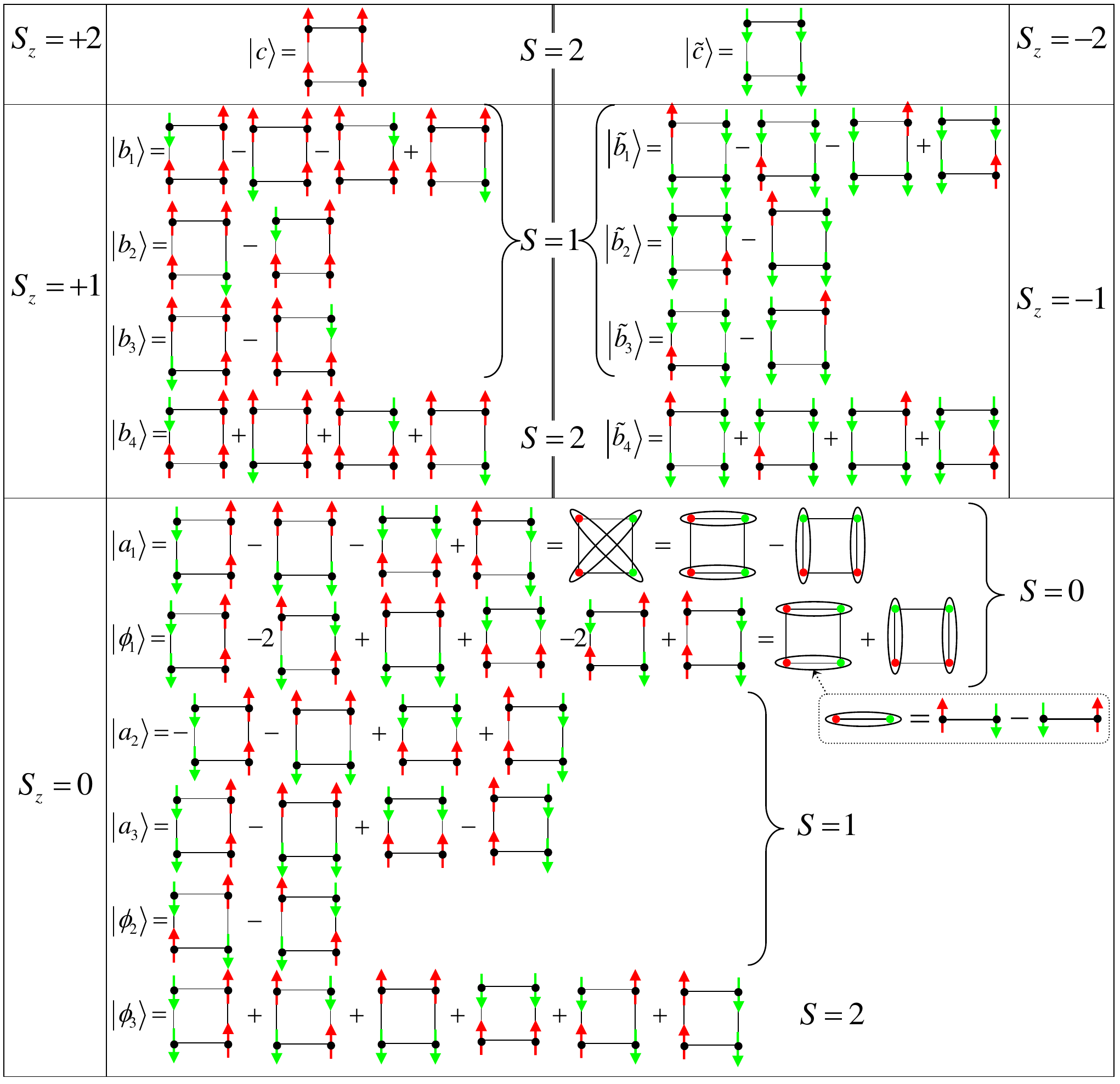}
\caption{Basis spin functions for the bond-operator technique in the case of four spins in the unit cell which has the form of a plaquette. Normalization factors are omitted for clarity. For each spin function, corresponding values are indicated of the total spin $S$ and its projection $S_z$. 
\label{statesfig}}
\end{figure}

We build the bond-operator formalism in the case of four spins in the unit cell using the basis presented graphically in Fig.~\ref{statesfig}. Bearing in mind the application in further discussion of this formalism to HAFs on square lattice, we choose the unit cell in the form of a plaquette. As soon as we derive the spin representation which can be used both in ordered and in paramagnetic phases, we choose states for the basis which are eigenfunctions of the total spin $S$ of the plaquette and its projection $S_z$ on quantized axis. Fifteen Bose-operators should be introduced which are labeled according to $S_z$ value of corresponding state (see Fig.~\ref{statesfig}):
\begin{align}
a_i^\dagger |0\rangle &= |a_i\rangle, 
&i&=1,2,3,4,5,\nonumber\\
b_i^\dagger |0\rangle &= |b_i\rangle, 
\quad 
\tilde b_i^\dagger |0\rangle = |\tilde b_i\rangle, 
&i&=1,2,3,4,\\
c^\dagger |0\rangle &= |c\rangle, 
\quad
\tilde c^\dagger |0\rangle = |\tilde c\rangle.\nonumber
\end{align}
Bosons $a$, $b$ ($\tilde b$), and $c$ ($\tilde c$) describe spin-0, spin-1, and spin-2 excitations, respectively. To be able to describe the N\'eel ordered phase, the wave function of the ground state $|0\rangle$ as well as $|a_4\rangle$ and $|a_5\rangle$ should be found as linear combinations of basis functions containing spin states with checkerboard motifs (i.e., $|\phi_{1,2,3}\rangle$ in Fig.~\ref{statesfig})
\footnote{
One could suggest to represent $|0\rangle$ and $|a_{1,2,3,4,5}\rangle$ states as linear combinations of all functions shown in Fig.~\ref{statesfig} from the sector with $S_z=0$. However, we have verified explicitly that such representation (containing five parameters instead of two in Eq.~\eqref{045}) leads to the same results in considered models. 
}
\begin{eqnarray}
\label{045}
|0\rangle &=& \cos\alpha \cos\beta |\phi_1\rangle +  \cos\alpha  \sin\beta |\phi_3\rangle -\sin\alpha |\phi_2\rangle,\nonumber\\
|a_4\rangle &=& \sin\alpha \cos\beta |\phi_1\rangle + \sin\alpha \sin\beta |\phi_3\rangle + \cos\alpha |\phi_2\rangle,\\
|a_5\rangle &=& - \sin\beta |\phi_1\rangle + \cos\beta |\phi_3\rangle.\nonumber
\end{eqnarray}
In particular, $\alpha=\beta=0$ in a HAF containing isolated plaquettes with exchange coupling between only nearest spins. 

We have realized the program of finding the spin representation which is proposed above for two spins in the unit cell: we have created an analog of Table~\ref{table1} and expressions similar to Eqs.~\eqref{trans} which have the same matrix elements. Then, as in Eqs.~\eqref{trans}, we have multiplied by $n$ constant terms and multiplied all terms linear in Bose-operators by the projector (cf.~Eq.~\eqref{proj})
\begin{equation}
\label{proj4}
P_j = \sqrt{
n 
- \sum_{i=1}^5a_{ij}^\dagger a_{ij}
- \sum_{i=1}^4 \left( b_{ij}^\dagger b_{ij} + \tilde b_{ij}^\dagger \tilde b_{ij} \right)
- c_j^\dagger c_j - \tilde c_j^\dagger \tilde c_j 
}.
\end{equation}
We have obtained as a result quite cumbersome expressions which are presented in Appendix~\ref{rep4}. It has been checked straightforwardly that the resultant expressions for spin components reproduce spin commutation algebra of operators ${\bf S}_{1j}$, ${\bf S}_{2j}$, ${\bf S}_{3j}$, and ${\bf S}_{4j}$ in $j$-th plaquette and that the Bose-analogs of operators 
$({\bf S}_{mj}{\bf S}_{nj})$, where $m,n=1,2,3,4$, commute with the Bose-analog of 
${\bf S}_{1j} + {\bf S}_{2j} + {\bf S}_{3j} + {\bf S}_{4j}$. 

\begin{figure}
\includegraphics[scale=0.4]{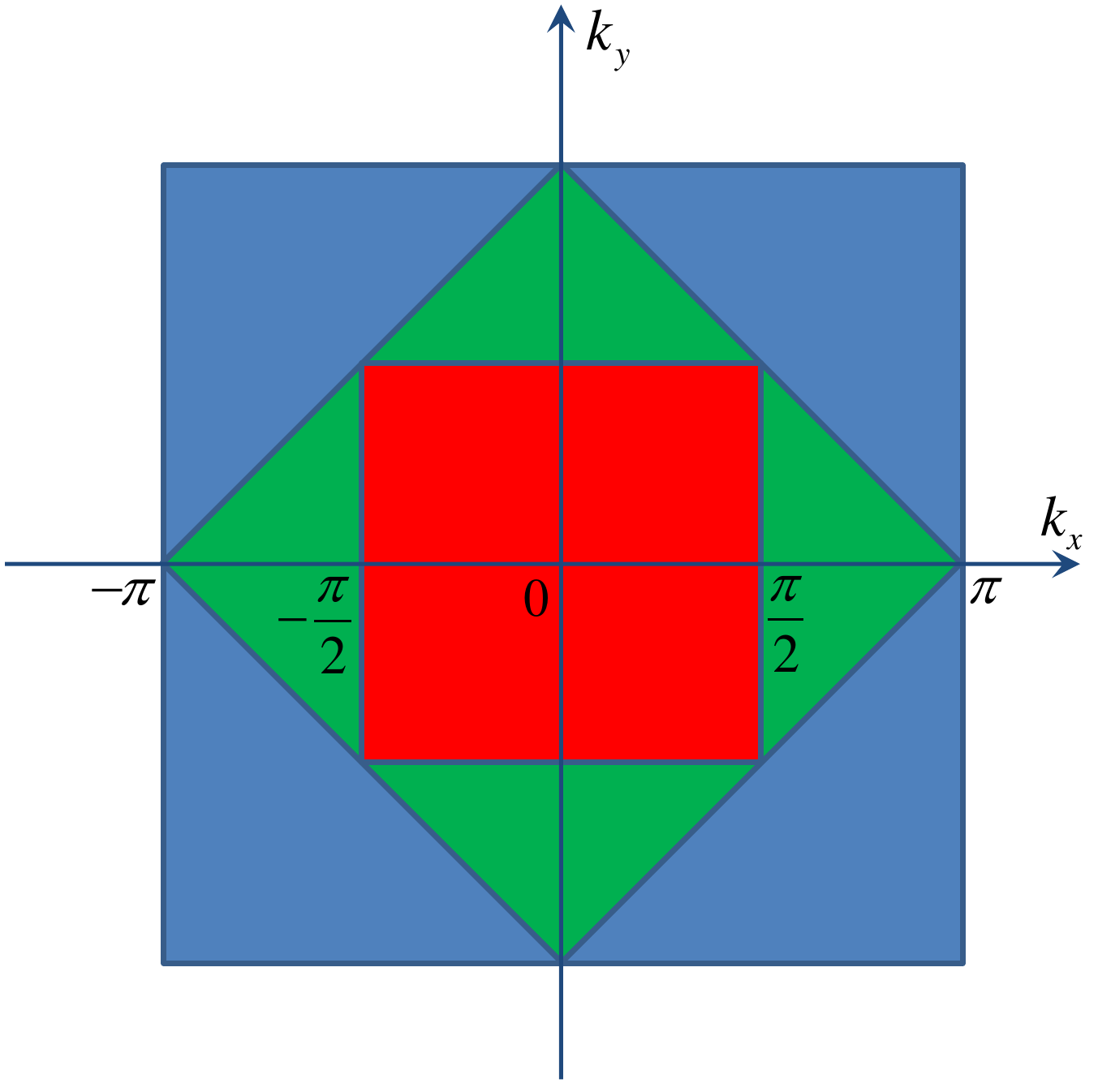}
\caption{The chemical and magnetic Brillouin zones (BZs) are presented (the largest and the middle squares, respectively) for the simple square lattice. The distance between nearest lattice sites is set to be equal to unity. The smallest (red) square and the green area are the first and the second BZs, correspondingly, in the case of four sites in the unit cell which has the form of a plaquette.
\label{bz}}
\end{figure}

It is seen that many new quasiparticles appear in the considered formalism as compared, e.g., with the conventional spin-wave theory or BOTs with two spins in the unit cell. One should bare in mind that momenta of quasiparticles in the proposed technique are restricted to the first Brillouin zone (BZ) which twice as little as the magnetic BZ (see Fig.~\ref{bz}). Then, four spin-1 bosons in the suggested technique should describe two magnons in the magnetic BZ. Spin-0 quasiparticles are from sector with $S_z=0$, where, as it is well known, bound states of two magnons and the amplitude (Higgs) mode live. Then, it is clear that two $a$-quasiparticles should correspond to the amplitude mode. We demonstrate below in detail by the example of HAF on simple square lattice how to identify magnons and the Higgs mode among spin-1 and spin-0 excitations, respectively, and how to restore their spectra in magnetic BZ from spectra of the introduced bosons found within the red region in Fig.~\ref{bz}. We find below that the rest four spin-1 elementary excitations (as well as spin-2 quasiparticles) describe high-energy excitations in spin-$\frac12$ HAF on square lattice. It is shown also that $a_1$-quasiparticle is a special elementary excitation which lies below the Higgs mode in the major part of BZ and which is purely singlet in paramagnetic phases. 

\section{Spin-$\frac12$ HAF on simple square lattice}
\label{ssl}

We apply now the formalism suggested in the previous section to spin-$\frac12$ HAF on simple square lattice whose Hamiltonian has the form
\begin{equation}
\label{ham}
{\cal H} = \sum_{\langle i,j \rangle}	{\bf S}_i{\bf S}_j,
\end{equation} 
where the exchange coupling constant is set to be equal to unity. We proceed in much the same manner as in the case of the square lattice bilayer. The difference is that all the derivations are lengthy and have to be done only on computer. 

\subsection{Ground-state energy and staggered magnetization}
\label{static4}

After the unit cell doubling in two directions and substitution of the Bose-analogs of spins operators presented in Appendix~\ref{rep4} to spin Hamiltonian \eqref{ham}, we obtain Eq.~\eqref{hbose}, where the first two terms have the form
\begin{eqnarray}
\label{e04}
\frac{{\cal E}}{N} &=&  
-\frac{n^2}{12} \left( 3 (4-\cos2 \alpha -\cos4 \alpha ) + 2 (11-2 \cos2 \alpha ) \cos^2\alpha \cos2 \beta - 4 \sqrt{2} \sin^2 2\alpha  \sin2 \beta \right), \\
\label{h14}
\frac{{\cal H}_1}{\sqrt N} &=& n^{3/2} 
\left(\frac{1}{4} (1-3 \cos2 \beta ) \sin2 \alpha +\frac{1}{6} \sin4 \alpha  \left(3+\cos2 \beta -2 \sqrt{2} \sin2 \beta \right)\right) 
\left(a_{4\bf0}+a_{4\bf0}^\dagger\right)\nonumber\\
&&{}+
n^{3/2} 
\left(\frac{4}{3} \sqrt{2} \cos\alpha \sin^2\alpha \cos2 \beta  + \frac{1}{6} (10 \cos\alpha - \cos3 \alpha ) \sin2 \beta \right) 
\left(a_{5\bf0}+a_{5\bf0}^\dagger\right),
\end{eqnarray}
and $N$ is the number of unit cells in the lattice. The rest terms in Eq.~\eqref{hbose} are quite lengthy and we do not present them here. ${\cal H}_1$ vanishes at values of $\alpha$ and $\beta$ which minimize $\cal E$. The staggered magnetization reads in the leading order in $1/n$ as
\begin{equation}
	M = n \frac{\sin2\alpha \left( \sqrt{2} \cos\beta - \sin\beta \right)}{2\sqrt{3}}.
\end{equation}
Taking into account first $1/n$ corrections to ${\cal H}_1$, to the ground-state energy $\cal E$, and to $M$, one obtains
\begin{eqnarray}
\label{alphass}
\alpha &=& 0.6486 -0.1081 \frac1n,\\
\label{betass}
\beta &=& -0.1879 +0.1396 \frac1n,\\
\label{eval4}
\frac{\cal E}{4N} &=& -0.5841n^2 - 0.0717n,\\
\label{s1zval4}
M &=&
\langle S_{1j}^z \rangle = 
-\langle S_{2j}^z \rangle 
=\langle S_{3j}^z \rangle 
=-\langle S_{4j}^z \rangle 
=0.4381n - 0.1367.
\end{eqnarray}
Eqs.~\eqref{eval4} and \eqref{s1zval4} give, correspondingly, $-0.656$ and $0.301$ at $n=1$ which are very close to values of $\approx-0.667$ and $\approx0.3$ obtained before by many methods \cite{monous}. Then, similar to $1/S$-expansion, first $1/n$ corrections give the main contribution to renormalization of the ground-state energy and the staggered magnetization. 
\footnote{Interestingly, one obtains nearly the same values of $-0.658$ and $0.303$ for the ground-state energy per spin and the staggered magnetization $M$, respectively, in the first order in $1/S$ at $S=1/2$. \cite{monous} The second-order correction in $1/S$ brings the ground-state energy to an excellent agreement with numerical results whereas further order corrections to $M$ are negligible.}

\subsection{Elementary excitations. Harmonic approximation.}
\label{excit4}

Before presenting spectra of quasiparticles in the first order in $1/n$, it is instructive to consider them in the harmonic approximation in special cases of weakly coupled plaquettes and in the Ising limit. This allows us to trace evolution of elementary excitations from the simple exactly-solvable limits to regimes with considerable quantum fluctuations. We also relate in this way some quasiparticles introduced in the suggested formalism with elementary excitations observed before by conventional methods.

\subsubsection{Isolated and interacting plaquettes}

Spin states presented in Fig.~\ref{statesfig} are eigenfunctions of an isolated plaquette, in which case the ground state $|0\rangle=|\phi_1\rangle$, $|a_4\rangle=|\phi_2\rangle$, and $|a_5\rangle=|\phi_3\rangle$ (i.e., $\alpha=\beta=0$ in Eq.~\eqref{045}). One obtains for the bilinear part of the Hamiltonian of HAF with zero inter-plaquette interaction
\begin{eqnarray}
\label{h2iso}
{\cal H}_2^{isol} &=& \sum_{\bf k}	
\left(
\left( b_{1\bf k}^\dagger b_{1\bf k} + \tilde b_{1\bf k}^\dagger \tilde b_{1\bf k} + a_{4\bf k}^\dagger a_{4\bf k}\right)\right.\nonumber\\
&&{}
+
2\left( b_{2\bf k}^\dagger b_{2\bf k} + \tilde b_{2\bf k}^\dagger \tilde b_{2\bf k} + b_{3\bf k}^\dagger b_{3\bf k} + \tilde b_{3\bf k}^\dagger \tilde b_{3\bf k} + a_{1\bf k}^\dagger a_{1\bf k} + a_{2\bf k}^\dagger a_{2\bf k} + a_{3\bf k}^\dagger a_{3\bf k} \right)\\
&&{}+\left.
3\left( c_{\bf k}^\dagger c_{\bf k} + \tilde c_{\bf k}^\dagger \tilde c_{\bf k} + b_{4\bf k}^\dagger b_{4\bf k} + \tilde b_{4\bf k}^\dagger \tilde b_{4\bf k}  + a_{5\bf k}^\dagger a_{5\bf k}\right)
\right).\nonumber
\end{eqnarray}
Then, three degenerate dispersionless branches arises in this limit. We trace the evolution of spectra by introducing the exchange coupling constant between nearest-neighbor spins from different plaquettes $\lambda$ ($\lambda=0$ and $\lambda=1$ correspond to fully isolated plaquettes and HAF on square lattice, respectively). The minimum of $\cal E$ is located at $\alpha=\beta=0$ for $\lambda<\lambda_c\approx0.375$ whereas $\alpha$, $\beta$, and $M$ become finite at $\lambda>\lambda_c$ signifying QPT to the ordered phase at $\lambda=\lambda_c$. 
\footnote{
In the first order in $1/n$, we have obtained $\lambda_c\approx0.452$ at $n=1$. This result is in reasonable agreement with previous considerations \cite{plaq1,plaq2,plaq3,plaq4,plaq5,plaq6} of this system by various methods which give for $\lambda_c$ values from the interval 0.47--0.6. $\lambda_c\approx0.55$ seems to be the most reliable result \cite{plaq2,plaq3,plaq6}.}

${\cal H}_2$ becomes very cumbersome at $\lambda\ne0$. It contains 55 and 95 terms at $\lambda<\lambda_c$ and $\lambda>\lambda_c$, respectively. In particular, at $\lambda>\lambda_c$, there are terms of the type $a_{m\bf k}^\dagger a_{n\bf k}$, $b_{m\bf k}^\dagger b_{n\bf k}$, and $\tilde b_{m\bf k}^\dagger \tilde b_{n\bf k}$ with $m\ne n$; terms $a_{m\bf k} a_{n-\bf k}$ (and $a_{m\bf k}^\dagger a_{n-\bf k}^\dagger$); and terms $b_{m\bf k} \tilde b_{n-\bf k}$ (and $b_{m\bf k}^\dagger \tilde b_{n-\bf k}^\dagger$). However, operators $c_{\bf k}$, $\tilde c_{\bf k}$, and $a_{1\bf k}$ enters in ${\cal H}_2$ only in combinations $c_{\bf k}^\dagger c_{\bf k}$, $\tilde c_{\bf k}^\dagger \tilde c_{\bf k}$, and $a_{1\bf k}^\dagger a_{1\bf k}$ at any $\lambda$. As a result, it is impossible to associate a spectrum branch with the introduced bosons at finite $\lambda$ (except for $c$, $\tilde c$, and $a_1$). Nevertheless, just for the purposes of better presentation and more convenient tracing of spectra evolution, we relate below a spectrum branch with the introduced bosons by considering residues of 15 Green's functions
\begin{equation}
\label{chi}
\chi_{AB}(\omega,{\bf k}) = 
i\int_0^\infty dt 
e^{i\omega t}	
\left\langle \left[ A_{\bf k}(t), B_{-\bf k}(0) \right] \right\rangle,
\end{equation}
where $A=B^\dagger$ runs over all $a$, $b$, $\tilde b$, $c$, and $\tilde c$ operators. We associate (roughly!) boson $A$ with a spectrum branch if the absolute value of the corresponding residue of $\chi_{AA^\dagger}(\omega,{\bf k})$ exceeds 0.15 at least on a half of the first BZ. Then, $A$ can be associated with more than one branch in this way.

As soon as operator of the inter-plaquette interaction commutes with the total spin, the classification is valid in the disordered phase of energy levels according to values of the total spin and its projection. Then, boson $a_1$ describes the only purely singlet quasiparticle in the singlet phase. We call these singlet excitations singlons for short. In the ordered phase, $a_1$-quasiparticles are not singlet because the classification of levels according to the total spin values breaks in the thermodynamical limit. \cite{auer,lecture1} To the best of our knowledge, no special name has been proposed for such quasiparticles in the ordered phase. Then, we call them below "singlons" (in quotes) in the ordered phase. 

It is seen from Eqs.~\eqref{s1jz}--\eqref{s4j+} that spectra of all spin-1 quasiparticles and all spin-0 ones (except for $a_1$!) appear in the ordered phase as poles of dynamical spin structure factors (DSSFs) $\chi_{+-}(\omega,{\bf k})$ and $\chi_{zz}(\omega,{\bf k})$ which are given by Eq.~\eqref{chi} with $A=S^+$, $B=S^-$ and  $A=B=S^z$, respectively ($\chi_{+-}(\omega,{\bf k})$ and $\chi_{zz}(\omega,{\bf k})$ contain in the leading order in $1/n$ Green's functions of $b$ ($\tilde b$) and $a_{2,3,4,5}$ operators, correspondingly). Spin operators read in our terms as $S^\gamma_{\bf k} = S^\gamma_{1\bf k} + e^{-ik_y/2}S^\gamma_{2\bf k} + e^{-i(k_x+k_y)/2} S^\gamma_{3\bf k} + e^{-ik_x/2}S^\gamma_{4\bf k}$, where the double distance between nearest spins is set to be equal to unity and spins in the unit cell are enumerated clockwise starting from its left lower corner. To probe $a_1$-quasiparticles, one has to consider many-spin correlators: e.g., the bond-bond correlator given by Eq.~\eqref{chi} with $A_{\bf k}=B_{\bf k}=\sum_i e^{-i({\bf k}{\bf R}_i)} ({\bf S}_{{\bf R}_i} {\bf S}_{{\bf R}_i+{\bf r}_y})$, where ${\bf r}_y$ is a vector connecting two nearest lattice sites. This correlator contains also poles corresponding to other spin-0 branches. It is also shown below that the Raman spectrum is related in the leading order in $1/n$ with the imaginary part of $\chi_{a_1a_1^\dagger}(\omega,{\bf 0})$.

\begin{figure}
\includegraphics[scale=0.7]{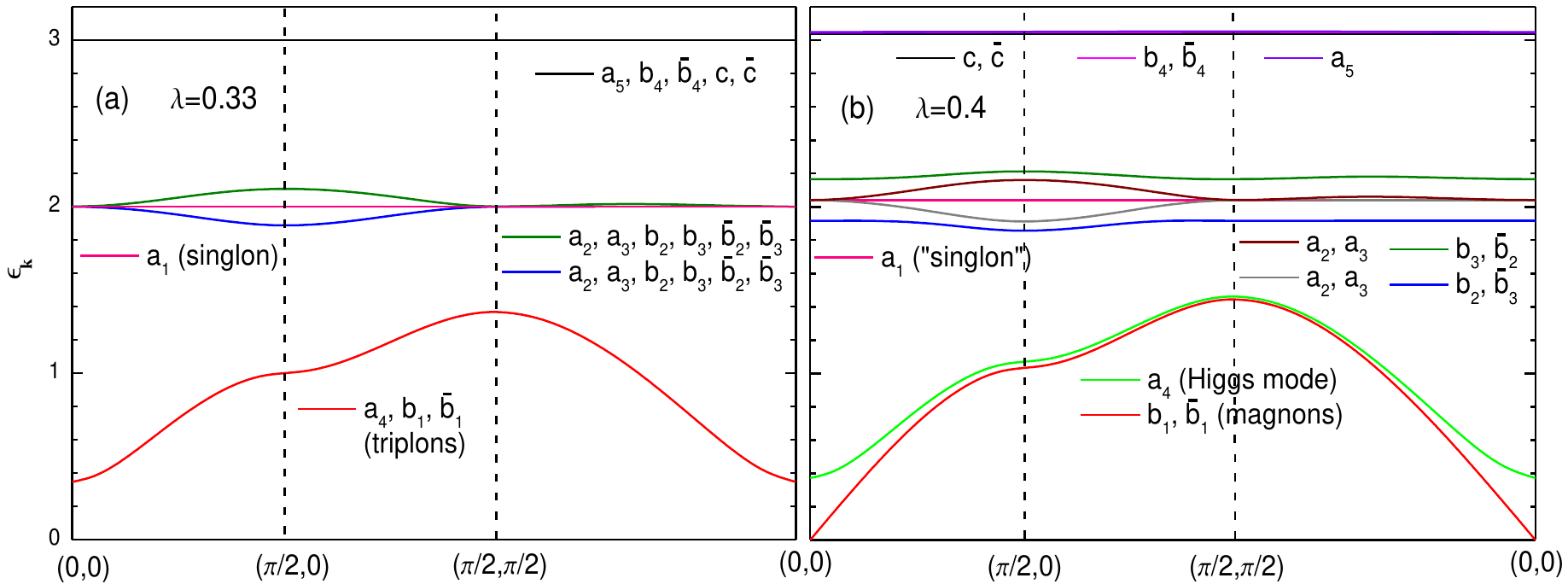}
\includegraphics[scale=0.65]{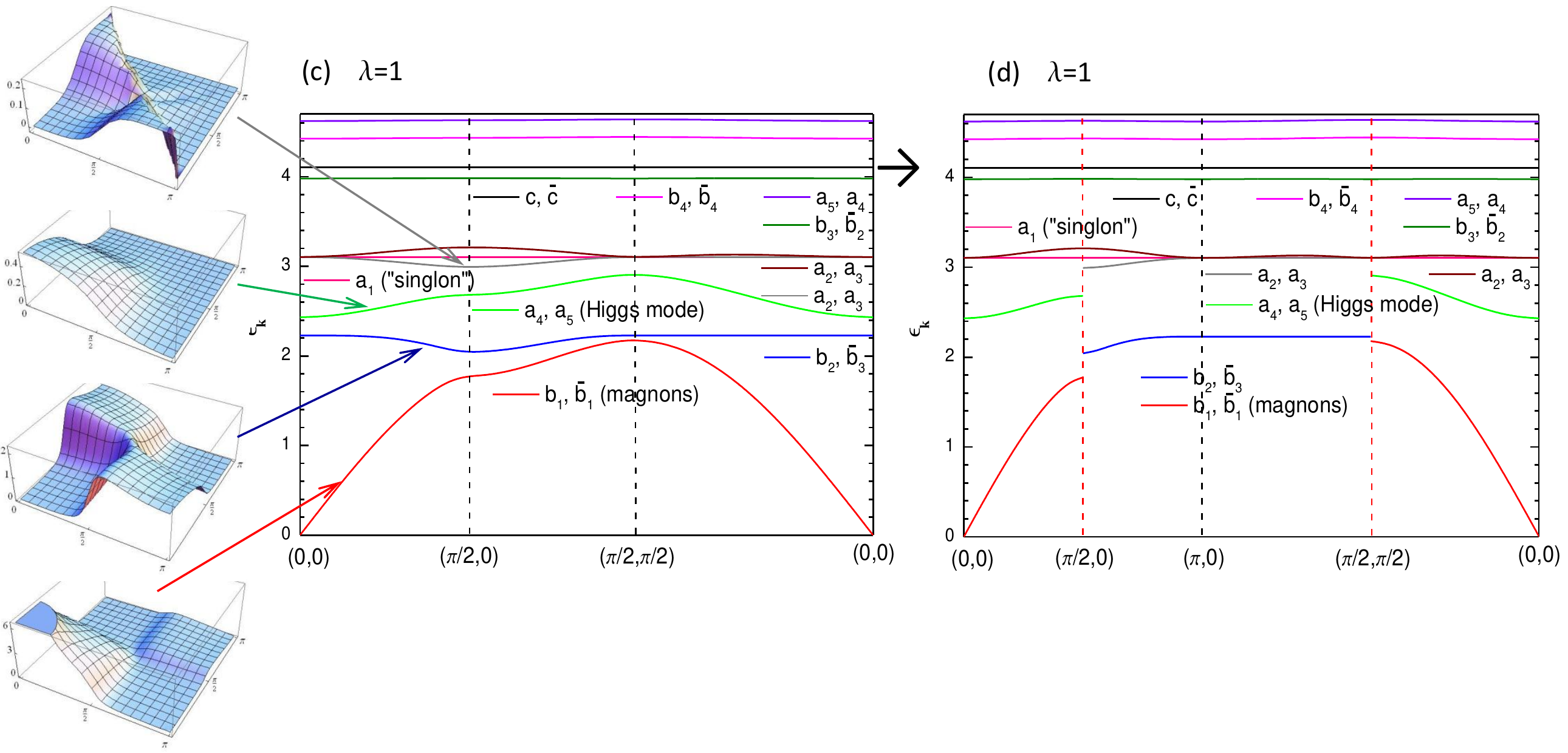}
\caption{Spectra of elementary excitations obtained using the suggested formalism in the harmonic approximation for selected values of parameter $\lambda$ which controls the strength of inter-plaquette coupling ($\lambda=0$ and $\lambda=1$ correspond to fully isolated plaquettes and HAF on square lattice, respectively). Each branch of excitations is associated with introduced bosons according to values of residues of the corresponding Green's function \eqref{chi} as it is explained in the text. Panels (a) and (b) describe the neighborhood of the QCP from the paramagnetic phase to the ordered one. Insets in panel (c) show residues of dynamical spin structure factors $\chi_{+-}$ (lower two insets) and $\chi_{zz}$ (upper two insets) corresponding to pointed branches. (d) Spectra presented in panel (c) but drawn in the magnetic BZ (see Fig.~\ref{bz}) as it is explained in the text.
\label{isolfig}}
\end{figure}

Spectra of quasiparticles in the harmonic approximation are shown in Fig.~\ref{isolfig} for selected values of $\lambda$. There are five different spectrum branches at $0<\lambda\le\lambda_c$ (see Fig.~\ref{isolfig}(a)). The lower branch is triply degenerate and it corresponds to the well-known triplons whose spectrum softens at $\lambda=\lambda_c$ and it splits at $\lambda\agt\lambda_c$. The branch characterized by $S_z=0$ detaches from the doubly degenerate branch of spin-1 excitations (magnons) forming the Higgs (amplitude) mode at $\lambda>\lambda_c$ (see Fig.~\ref{isolfig}(b)). 

Spectra are presented in Fig.~\ref{isolfig}(c) at $\lambda=1$. Notice that the lattice symmetry is restored at $\lambda=1$ and one has to recover somehow within the considered formalism the conventional picture of elementary excitations of HAF with two degenerate magnon and one amplitude modes in the magnetic BZ (see Fig.~\ref{bz}). It is easy to see from Fig.~\ref{isolfig}(c) that a simple extension of obtained spectra to the green area in Fig.~\ref{bz} (which is the second BZ in this case) would lead to low-energy spin-1 excitations in the magnetic BZ having zero energy at ${\bf k}=(\pi,0)$ that would contradict the conventional wisdom about magnons. The common picture can be restored by consideration of observable quantities (e.g., DSSFs). Let us consider first the transverse DSSF $\chi_{+-}(\omega,{\bf k})$ in the leading order in $1/n$ (i.e., we take into account only linear in Bose-operators terms in Eqs.~\eqref{s1j+}--\eqref{s4j+}). Then, $\chi_{+-}(\omega,{\bf k})$ contains only Green's functions of $b$- and $\tilde b$-operators. Graphics of its residues (shifted by ${\bf k}_0=(\pi,\pi)$ for convenience) corresponding to two lower spin-1 branches are shown in two lower insets of Fig.~\ref{isolfig}(c). It is seen that the residue corresponding to the lower spin-1 branch is finite inside the red area in Fig.~\ref{bz} and it drops rapidly upon going deep into the green area (in particular, it is exactly zero at ${\bf k}=(\pi,0)$ in the considered harmonic approximation). The situation with the second spin-1 branch is inverse: the residue is finite within the green area and it drops rapidly to zero inside the red area. As it is seen from two upper insets of Fig.~\ref{isolfig}(c), similar situation arises in the case of two lowest branches of spin-0 excitations upon consideration of $\chi_{zz}(\omega,{\bf k}+{\bf k}_0)$. Residues of other spin-0 and spin-1 excitations do not show similar rapid reductions inside red or green areas. We draw in Fig.~\ref{isolfig}(d) the obtained spectra in the magnetic BZ not showing branches in the red and green areas with drastically reduced corresponding residues of DSSFs. It is shown below that the gaps between red and blue (green and gray) curves on borders of the red and the green areas are reduced in the first order in $1/n$ so that the curves in these two couples look more like continuations of each other. However, the gaps in the magnon and the amplitude mode spectra do not disappear completely in the first order in $1/n$.

\subsubsection{Ising-type anisotropy and Ising limit}
\label{izing}

It is instructive also to consider within the suggested formalism HAF with Ising-type anisotropy 
\begin{equation}
	{\cal H}^{Ising} = \sum_{\langle i,j \rangle}	
	\left(S_i^z S_j^z + \frac{A}{2}\left(S_i^+ S_j^-+S_i^- S_j^+\right)\right),
\end{equation}
where $0\le A<1$. Of particular interest is the exactly solvable Ising limit ($A=0$) in which case one obtains $\alpha=\pi/4$, $\tan 2\beta=-2\sqrt2$, $M=0.5$, ${\cal H}_{2i+1}=0$, where $i$ is integer, and
\begin{eqnarray}
\label{h2ising}
{\cal H}_2^{Ising} &=& \sum_{\bf k}	
\left(
2\left( b_{2\bf k}^\dagger b_{2\bf k} + {\mathfrak b}_{1\bf k}^\dagger {\mathfrak b}_{1\bf k} + \tilde {\mathfrak b}_{4\bf k}^\dagger \tilde {\mathfrak b}_{4\bf k} + \tilde b_{3\bf k}^\dagger \tilde b_{3\bf k} \right)\right.\nonumber\\
&&{}+
3\left( a_{1\bf k}^\dagger a_{1\bf k} + a_{2\bf k}^\dagger a_{2\bf k} + a_{3\bf k}^\dagger a_{3\bf k} + a_{5\bf k}^\dagger a_{5\bf k} \right)\\
&&{}+\left.
4\left( c_{\bf k}^\dagger c_{\bf k} + \tilde c_{\bf k}^\dagger \tilde c_{\bf k} + {\mathfrak b}_{4\bf k}^\dagger {\mathfrak b}_{4\bf k} + b_{3\bf k}^\dagger b_{3\bf k} + \tilde b_{2\bf k}^\dagger \tilde b_{2\bf k} + \tilde {\mathfrak b}_{1\bf k}^\dagger \tilde {\mathfrak b}_{1\bf k}  + a_{4\bf k}^\dagger a_{4\bf k}\right)
\right),\nonumber
\end{eqnarray}
where the following Bose-operators are introduced: ${\mathfrak b}_{1\bf k} = (b_{1\bf k} + b_{4\bf k})/\sqrt2$, ${\mathfrak b}_{4\bf k} = (b_{1\bf k} - b_{4\bf k})/\sqrt2$,
$\tilde{\mathfrak b}_{1\bf k} = (\tilde b_{1\bf k} + \tilde b_{4\bf k})/\sqrt2$, and
$\tilde{\mathfrak b}_{4\bf k} = (\tilde b_{1\bf k} - \tilde b_{4\bf k})/\sqrt2$. It can be shown using the spin representation presented in Appendix~\ref{rep4} that there are no $1/n$-corrections to spectra of quasiparticles because all corresponding diagrams contain contours which can be walked around while moving by arrows of Green's functions (integrals over frequencies in such diagrams give zero). 
\footnote{The existence of only this kind of diagrams stems from the fact that each term in ${\cal H}_{2i}$ contains $i$ operators of creation and $i$ operators of annihilation.} 
Then, we observe in magnetic BZ two degenerate spin-1 modes (magnons) with energy 2 (the well-known result \cite{oguchi}) and four spin-0 excitations within the red area in Fig.~\ref{bz} having energy 3 (see Eq.~\eqref{h2ising}). It is well known that there are four two-magnon bound states with energy 3 within the magnetic BZ in the Ising antiferromagnet. \cite{oguchi} Then, four spin-0 modes observed using the proposed technique correspond to the conventional two-magnon bound states. Consideration of $\chi_{zz}(\omega,{\bf k}+{\bf k}_0)$ at $0<A<1$ similar to that presented above shows that two of four lower spin-0 modes are continuation of each other in the red and green areas in Fig.~\ref{bz} (one obtains pictures similar to two upper insets in Fig.~\ref{isolfig}(c)). We believe that the rest two-magnon bound states arise in our formalism as bound states of two spin-1 excitations. However, a detailed consideration of this point is out of the scope of the present paper. 

Notice that the Higgs mode (green and grey curves in Figs.~\ref{isolfig}(c) and \ref{isolfig}(d)) as well as "singlons" stem from the two-magnon bound state modes in the considered HAF with the Ising anisotropy. We observe that they dive into the two-magnon continuum at $A\approx0.8$ in agreement with previous considerations \cite{oguchi,hamer2}.

\subsection{Elementary excitations. Renormalized spectra.}
\label{excithaf}

\begin{figure}
\includegraphics[scale=0.7]{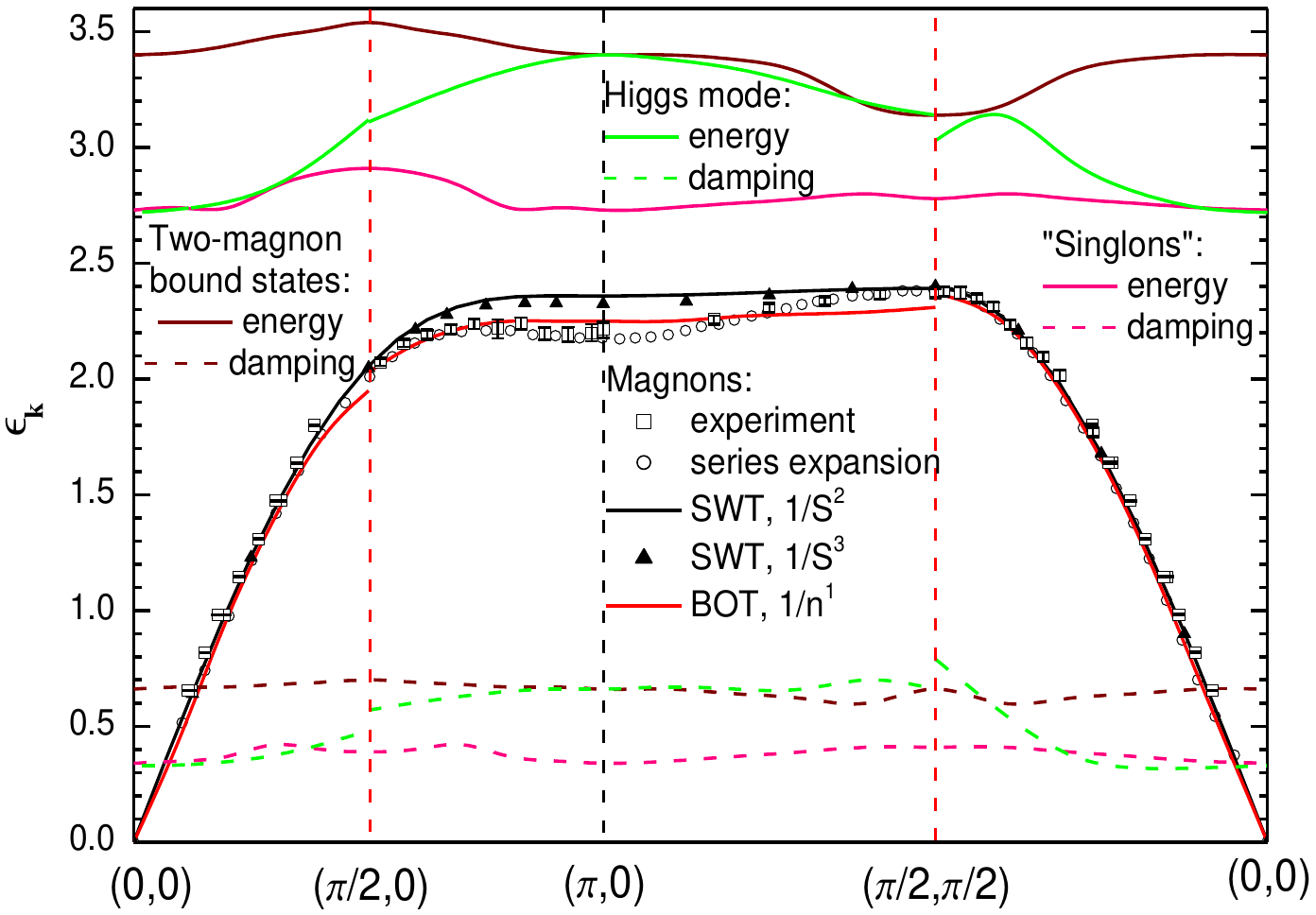}
\caption{Spectra of low-energy elementary excitations in spin-$\frac12$ HAF on square lattice found using the proposed bond-operator technique (BOT) in the first order in $1/n$. Also shown are magnon spectra obtained by series expansion around the Ising limit \cite{ser}, within the spin-wave theory (SWT) in the second \cite{igar,igar2} and in the third \cite{syromyat} orders in $1/S$, and neutron scattering experiment in CFTD \cite{chris1,piazza}. Borders of the first BZ with four spins in the unit cell are shown by red vertical lines (see Fig.~\ref{bz}). 
\label{spec_ss}}
\end{figure}

Spectra of low-energy elementary excitations found in the first order in $1/n$ are shown in Fig.~\ref{spec_ss} (cf.\ Fig.~\ref{isolfig}(d)). It is seen from Fig.~\ref{spec_ss} that magnon spectrum obtained within our technique is in good quantitative agreement with experiment in CFTD (the worse agreement is near borders between green and red areas in Fig.~\ref{bz}). In particular, notice a good quantitative agreement near ${\bf k}=(\pi,0)$, where $1/S$-expansion shows slow convergence pointed out in Ref.~\cite{syromyat}.

The experimental data in CFTD are described perfectly within two different theoretical approaches suggested in Refs.~\cite{piazza,spinon} and Refs.~\cite{pow1,pow2}. It is argued in  Refs.~\cite{pow1,pow2} that the deep in the magnon spectrum around ${\bf k}=(\pi,0)$ is due to the magnon attraction stimulated by strong magnon-Higgs scattering. Within our approach, the magnon-Higgs interaction comes from the diagram shown in Fig.~\ref{diag}(c), where one intrinsic line stands for the magnon Green's function and another line corresponds to Green's functions of $a_{2,3,4,5}$ operators. However, the magnon spectrum at ${\bf k}=(\pi,0)$ is not practically renormalized by $1/n$ corrections: $\epsilon_{(\pi,0)}\approx2.23$ and 2.25 in the harmonic approximation and in the first order in $1/n$, respectively (values of corrections from $\alpha$ and $\beta$ renormalization, and from diagrams shown in Figs.~\ref{diag}(b), and \ref{diag}(c) are $-0.08$, $0.91$, and $-0.81$, correspondingly). Then, our results do not support clearly the magnon attraction picture as a source of the spectrum anomaly near ${\bf k}=(\pi,0)$. As for previous explanations of this anomaly as a result of deconfinement \cite{piazza} or "partial deconfinement" \cite{spinon} of magnons into two spinons near ${\bf k}=(\pi,0)$, our approach is not intended to treat magnons in this way. Then, we cannot confirm using our results neither of the physical pictures suggested so far for the magnon spectrum anomaly near ${\bf k}=(\pi,0)$. A comprehensive consideration of the neighborhood of ${\bf k}=(\pi,0)$ requires also time-consuming calculations of DSSFs within the suggested formalism which will be carried out elsewhere.

We suggest in our recent papers \cite{singlon,wej1j2} an approach for description of low-energy singlet sector of spin-$\frac12$ HAFs. In particular, a spectrum of low-energy singlet excitations can be found by this technique. While it is naturally to expect that this approach is suitable for disordered phases with singlet ground states, we try to apply it in Ref.~\cite{singlon} to HAF on simple square lattice taking into account that all excitations in the ordered phase can be classified according to the spin value before proceeding to thermodynamical limit \cite{auer,lecture1}. We obtain in Ref.~\cite{singlon} that the spectrum of singlet excitations lies below the magnon spectrum around ${\bf k}=(\pi,0)$. Most likely, this result is an artifact related to the fact that we go in Ref.~\cite{singlon} beyond the method applicability. This conclusion is supported by consideration of the Raman intensity in the next section, where we show that the position of the peak obtained experimentally in layered cuprates coincides with the "singlon" energy at ${\bf k=0}$ (spectra are equivalent of $a_1$-boson at ${\bf k=0}$ and at ${\bf k}=(\pi,0)$). Besides, it will be shown in our forthcoming paper \cite{weunp} that singlon spectra in the disordered phase of $J_1$--$J_2$ HAF on square lattice found within the first order in $1/n$ are in excellent agreement with those obtained in Ref.~\cite{wej1j2}.

It is seen also from Fig.~\ref{spec_ss} that there are moderately damped spin-0 excitations above the magnon branch the lower of which are the amplitude mode and "singlons". Remarkably, "singlons" lie below the Higgs mode almost in the whole BZ. As it is pointed out above, "singlons" cannot be detected explicitly via DSSFs. Only many-spin correlators can contain a contribution from the Green's function of $a_1$-quasiparticles. We show now that Raman scattering in $B_{1g}$ geometry probes these excitations with ${\bf k}={\bf 0}$.

\subsection{Raman spectrum}
\label{ramansec}

The standard theory of Raman scattering is based on an effective Loudon-Fleury Hamiltonian for the interaction of light with spin degrees of freedom which has the form
$
H_{LF}=
\sum_{\langle q,m \rangle}	\left({\bf e}_i {\bf R}_{qm} \right) \left({\bf e}_f {\bf R}_{qm} \right)
\left({\bf S}_q{\bf S}_m\right)
$,
where a common factor is omitted in the right-hand side, ${\bf e}_i$ and ${\bf e}_f$ are polarization vectors of incoming and outgoing photons, and ${\bf R}_{qm}$ is a vector connecting nearest-neighbor sites. \cite{lframan} This theory is expected to work well when energies of incoming and outgoing photons are considerably smaller than the gap between conduction and valence bands. Much attention has been paid previously to the Raman scattering in the so-called $B_{1g}$ symmetry in which case ${\bf e}_i$ is directed along a diagonal of a square, ${\bf e}_f\perp{\bf e}_i$, and the intensity of light is proportional to the imaginary part of susceptibility \eqref{chi}, where $A_{\bf k}$ and $B_{-\bf k}$ should be replaced by
\begin{equation}
\label{b1g}
H_{LF}^{B_{1g}} =
\sum_{\langle q,m \rangle}	{\bf S}_q{\bf S}_m 
-
\sum_{\langle\langle q,m \rangle\rangle}	{\bf S}_q{\bf S}_m,
\end{equation}
$\langle q,m \rangle$ and $\langle\langle q,m \rangle\rangle$ denote nearest neighbor spins along $x$ and $y$ directions, respectively. 

\begin{figure}
\includegraphics[scale=0.63]{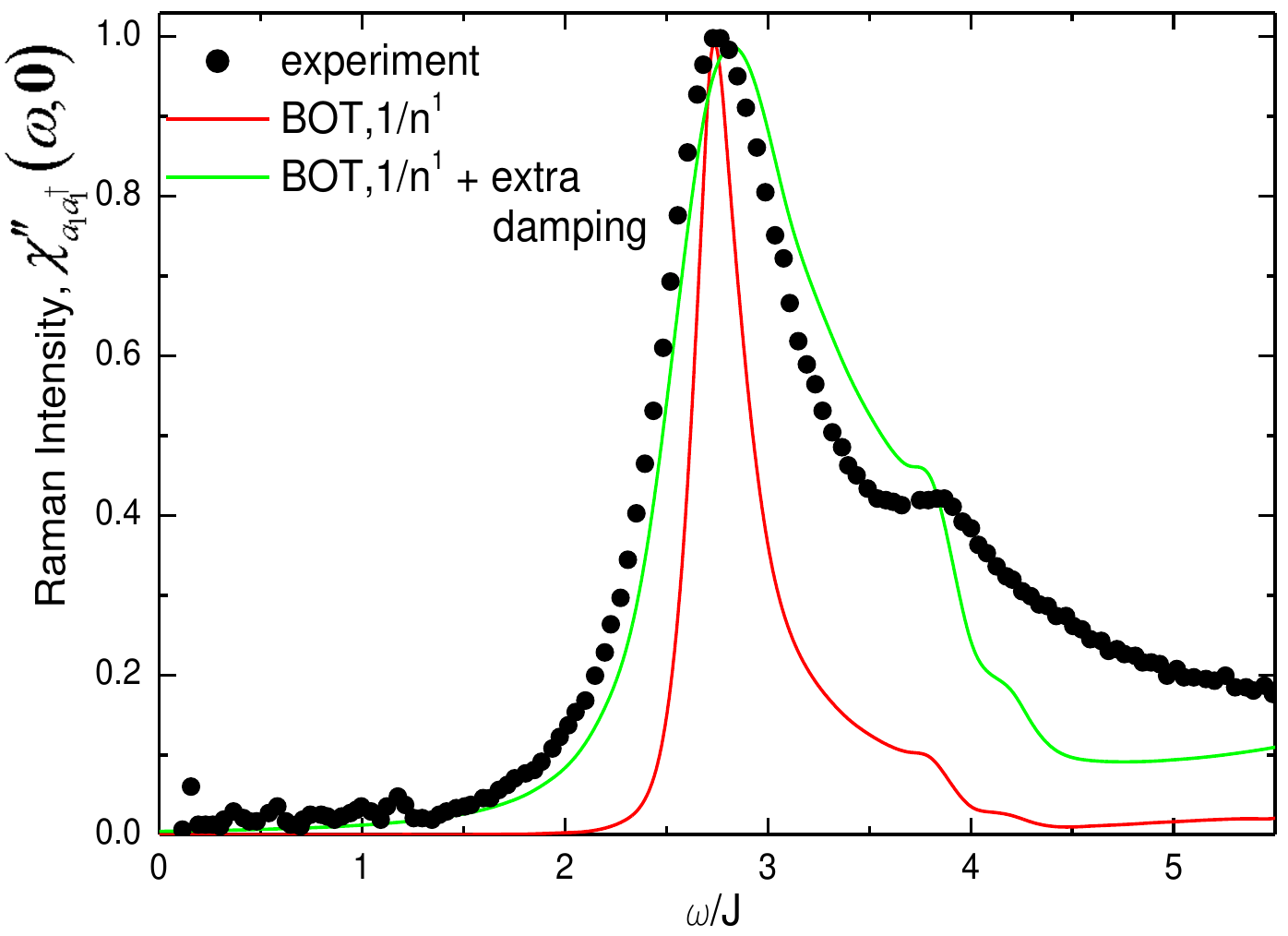}
\caption{Raman spectrum obtained experimentally in $\rm La_2CuO_4$ (taken from Ref.~\cite{ramanexp1}) and using the suggested bond-operator theory (BOT). In the leading order in $1/n$, the Raman intensity is given by the imaginary part of the "singlon" susceptibility $\chi''_{a_1a_1^\dagger}(\omega,{\bf 0})$ (see Eqs.~\eqref{i} and \eqref{chia1}). Experimental data are presented for the exchange coupling constant $J=147$~meV which is close to the value 143(2)~meV obtained from inelastic neutron scattering \cite{lco}. The green line is the result of calculation of $\chi''_{a_1a_1^\dagger}(\omega,{\bf 0})$ in the first order in $1/n$ with extra damping of $0.4J$ of all high-energy quasiparticles (see the text). Each set of data is multiplied by a factor to make the peak height to be equal to unity.
\label{raman_fig}}
\end{figure}

It is well known that in square-lattice HAF the $B_{1g}$ Raman spectrum has a broad asymmetric peak (referred to in the literature as "two-magnon" peak) at $\omega\approx3$ which has been attributed to scattering from magnon pairs with opposite momenta. \cite{elliott,raman1,raman2,raman3,raman4} In particular, this picture has been obtained in the insulating parent compounds of high-$T_c$ superconductors. \cite{ramanexp1,ramanexp2} There is also a shoulder-like structure at $\omega\approx4$ in $\rm La_2CuO_4$ (see Fig.~\ref{raman_fig}). Within the spin-wave theory, the $B_{1g}$ scattering is dominated by two-magnon excitations which give a peak around $\omega\approx3.3$ as a result of ladder diagrams summation. \cite{raman1,raman2,raman3,raman4} However, the peak form and the shoulder-like feature appearing in some compounds have not been explained within the spin-wave theory. It has been argued recently by expressing the problem in terms of an effective $O(3)$-model that the Raman spectrum contains a two-magnon and a two-Higgs contribution. \cite{weidinger2015} It is demonstrated in Ref.~\cite{weidinger2015} that the latter can be responsible for the shoulder-like anomaly in $\rm La_2CuO_4$.

Within our formalism, one obtains in the leading order in $1/n$ from Eqs.~\eqref{b1g}, \eqref{s1jz}--\eqref{s4j+}, and \eqref{b1gbond}
\begin{equation}
\label{hb1g}
	H_{LF}^{B_{1g}} = \sqrt{3nN} \cos \alpha  \cos \beta  \left( a_{1\bf 0}+a_{1\bf 0}^\dagger \right).
\end{equation}
Then, the Raman intensity has the form in the leading order in $1/n$
\begin{equation}
\label{i}
I(\omega) = 3nN \cos^2\alpha \cos^2\beta\left(\chi''_{a_1a_1^\dagger}(\omega,{\bf 0}) + \chi''_{a_1^\dagger a_1}(\omega,{\bf 0})\right),
\end{equation}
where $\chi''_{a_1a_1^\dagger}(\omega,{\bf 0})$ is the imaginary part of susceptibility \eqref{chi} with $A=B^\dagger=a_1$. Contribution of the second term in brackets in Eq.~\eqref{i} is negligibly small compared to that from the first term which reads as
\begin{equation}
\label{chia1}
	\chi''_{a_1a_1^\dagger}(\omega,{\bf 0}) = -{\rm Im}\left(\frac{1}{\omega-\epsilon_{0\bf 0}^{(a_1)}-\Sigma_{a_1}(\omega,{\bf 0})}\right),
\end{equation}
where $\epsilon_{0\bf k}^{(a_1)}\approx3.1$ is the bare spectrum of "singlons". We obtain after calculation of the self-energy part $\Sigma_{a_1}(\omega,{\bf 0})$ in the first order in $1/n$ that Eq.~\eqref{chia1} shows an asymmetric peak at $\omega\approx2.74$ corresponding to renormalized "singlon" energy at $\bf k=0$ (see Fig.~\ref{spec_ss}) and a shoulder extending up to $\omega\approx4$ (see Fig.~\ref{raman_fig}). It is seen from Fig.~\ref{raman_fig} that the position of the peak in $\rm La_2CuO_4$ (taken as an example) is reproduced quite accurately whereas the peak width is underestimated. The spectral weight of the peak is equal to $3nN\pi \cos^2\alpha \cos^2\beta \approx 5.6Nn$ in the leading order in $1/n$ (see Eqs.~\eqref{alphass}, \eqref{betass}, \eqref{i}, and \eqref{chia1}). This value is comparable at $n=1$ with the spectral weight of $\approx4.4N$ (calculated using Eqs.~(3.23) or (3.28) of Ref.~\cite{raman3}) of the "two-magnon" peak obtained at $\omega\approx3.3$ within the spin-wave formalism. Notice also that the decay of singlons into two spin-1 excitations makes the main contribution  to the imaginary part of $\Sigma_{a_1}(\omega,{\bf 0})$ at $\omega=3\div4$ (i.e., in the shoulder region).

\begin{figure}
\includegraphics[scale=0.08]{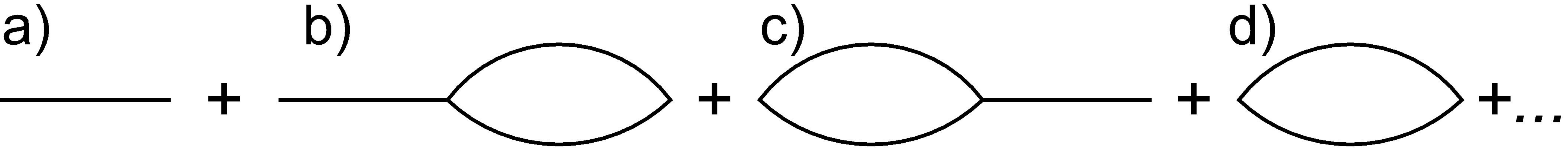}
\caption{Diagrams in the first few orders in $1/n$ for spin-spin and "scalar" correlators considered within the suggested bond-operator formalism.
\label{chifig}}
\end{figure}

Indeed, one has to consider the Raman intensity in further orders in $1/n$, where, in particular, diagrams appear describing two-spin-1 and two-spin-0 contributions (see Eqs.~\eqref{b1g}, \eqref{s1jz}--\eqref{s4j+}, \eqref{b1gbond}, and Fig.~\ref{chifig}). The corresponding analysis requires quite time-consuming calculations which will be carried out in future. Here, we present only the result of non-rigorous attempt to go beyond the first order in $1/n$ by taking into account the most pronounced renormalization of bare spectra. The latter is the finite damping of all elementary excitations except for magnons arising in the first order in $1/n$ (renormalization of quasiparticles energies does not exceed 20\%). To take into account the quasiparticles damping phenomenologically, we repeat the calculation of the self-energy part $\Sigma_{a_1}(\omega,{\bf k})$ in Eq.~\eqref{chia1} in the first order in $1/n$ adding "by hand" $0.4i$ with proper signs to all poles of Green's functions except for those corresponding to magnons (cf.~Fig.~\ref{spec_ss}). The result for the Raman intensity is presented in Fig.~\ref{raman_fig} by green line. The peak position and its spectral weight do not practically change while its width triples. 

\section{Summary and conclusion}
\label{conc}

In this paper, we present a bond-operator theory (BOT) for description both magnetically ordered phases and paramagnetic phases with singlet ground states in spin-$\frac12$ magnetic systems. This technique provides a regular expansion of physical quantities in powers of $1/n$, where $n$ is the maximum number of bosons which can occupy a unit cell (physical results indeed correspond only to $n=1$). Two variants of BOT are suggested: for two and for four spins in the unit cell. To probe the formalism, we consider first a paradigmatic model with two spins in the unit cell, spin-$\frac12$ HAF on square lattice bilayer, which has been discussed before by many other methods. We show that the ground-state energy $\cal E$, the staggered magnetization $M$, and quasiparticles spectra found within the first order in $1/n$ are in good {\it quantitative} agreement with previous results both in paramagnetic and in ordered phases not very close to QCP between the two. 

By doubling the unit cell in two directions, we discuss spin-$\frac12$ HAF on square lattice using the suggested BOT with four spins in the unit cell. We identify spin-1 magnon and spin-0 amplitude (Higgs) modes among fifteen spin-2, spin-1, and spin-0 elementary excitations. $\cal E$, and $M$ found in the first order in $1/n$ are, respectively, in good and in excellent {\it quantitative} agreement with previous numerical and experimental results. Magnon spectrum calculated in the first order in $1/n$ is also in good quantitative agreement with previous experimental and numerical results even around ${\bf k}=(\pi,0)$, where a deep in the spectrum was found not described quantitatively by standard analytical approaches including $1/S$-expansion.

We find a special spin-0 quasiparticle which is purely singlet (singlon) in paramagnetic phase, which has not been discussed widely so far in the ordered phases, and which lie below the Higgs mode in the ordered phase of spin-$\frac12$ HAF in the most part of the Brillouin zone. We call it "singlon" (in quotes) in the ordered state as it is no more singlet upon the  breaking of the continuous symmetry. By considering HAF with Ising-type anisotropy, we show that both Higgs and "singlon" modes stem from two-magnon bound states which merge with two-magnon continuum not far from the isotropic limit. We demonstrate that "singlons" do not appear explicitly in spin susceptibilities but they become visible in scalar correlators one of which describes the Raman intensity in $B_{1g}$ symmetry. We show that the latter is expressed in the leading order in $1/n$ via the "singlon" Green's function at zero momentum which shows an asymmetric peak. The position of this peak coincides with the position of the "two-magnon" peak observed experimentally in, e.g., layered cuprates. The spectral weight of this peak is comparable with that of the "two-magnon" peak obtained before within $1/S$-expansion in the ladder approximation. However, an analysis is needed in further orders in $1/n$ to describe the experimental data in every detail which will be performed elsewhere.

The suggested BOTs appear as efficient (although quite cumbersome) techniques allowing to discuss not only the well-known elementary excitations (magnons and triplons) but also those which arise in conventional techniques as poles of many-particle vertexes (the amplitude mode, singlons, two-magnon or two-triplon bound states).

\begin{acknowledgments}

We thank N.B.~Christensen, D.\ Joshi, and O.P.\ Sushkov for exchange of data and useful discussions. This work is supported by Foundation for the advancement of theoretical physics and mathematics "BASIS".

\end{acknowledgments}

\appendix

\section{Spin representation for four spins in the unit cell}
\label{rep4}

We present in this appendix Bose-analogs of spin operators in the case of four spins in the unit cell having the form of a plaquette. All expressions have been derived as it is explained in the main text (see Sec.~\ref{method4}). Spins are enumerated in the plaquette clockwise starting from its left lower corner.
\begin{eqnarray}
\label{s1jz}
S_{1j}^z &=& -nu_1 \sin2\alpha
+
u_2 \cos\alpha \left(P_ja_{2j}+a_{2j}^\dagger P_j\right) 
+ u_1 \cos2\alpha \left(P_ja_{4j}+a_{4j}^\dagger P_j\right)
- u_2 \sin\alpha \left(P_ja_{5j}+a_{5j}^\dagger P_j\right)\\
&&+
u_1 \sin2\alpha \left(a_{1j}^\dagger a_{1j} + a_{2j}^\dagger a_{2j}+a_{3j}^\dagger a_{3j} +2 a_{4j}^\dagger a_{4j}+ a_{5j}^\dagger a_{5j}+ b_{3j}^\dagger b_{3j}+\tilde b_{3j}^\dagger\tilde b_{3j} \right)\nonumber\\
&&+ \left(\frac{1}{4}+u_1 \sin2\alpha\right) \left(b_{1j}^\dagger b_{1j} +b_{4j}^\dagger b_{4j} \right) + \left(\frac{1}{2}+u_1 \sin2\alpha\right) \left( b_{2j}^\dagger b_{2j}+ c_j^\dagger c_j \right)\nonumber\\
&&-\left(\frac{1}{4}-u_1 \sin2\alpha\right) \left( \tilde b_{1j}^\dagger\tilde b_{1j}+ \tilde b_{4j}^\dagger\tilde b_{4j}\right) - \left(\frac{1}{2}-u_1 \sin2\alpha\right) \left(\tilde b_{2j}^\dagger\tilde b_{2j} +\tilde c_j^\dagger\tilde c_j  \right)\nonumber\\
&&+\frac{1}{4} \left(4 u_2 \sin\alpha a_{2j}^\dagger a_{4j}-2 a_{1j}^\dagger a_{3j} - 4 u_1 a_{2j}^\dagger a_{5j}+4 u_2 \cos\alpha a_{4j}^\dagger a_{5j} + \sqrt{2} b_{1j}^\dagger b_{3j} + b_{1j}^\dagger b_{4j}\right.\nonumber\\
&&-\left.\sqrt{2} b_{3j}^\dagger b_{4j}-\sqrt{2} \tilde b_{1j}^\dagger \tilde b_{3j}-\tilde b_{1j}^\dagger \tilde b_{4j}+\sqrt{2} \tilde b_{3j}^\dagger \tilde b_{4j}+h.c.\right),\nonumber\\
S_{2j}^z &=& nu_1 \sin2\alpha
- u_2 \cos\alpha \left(P_ja_{3j}+a_{3j}^\dagger P_j\right)
-u_1 \cos2\alpha \left(P_ja_{4j}+a_{4j}^\dagger P_j\right)
+u_2 \sin\alpha \left(P_ja_{5j}+a_{5j}^\dagger P_j\right)\\
&&-u_1 \sin2\alpha \left(a_{1j}^\dagger a_{1j} + a_{2j}^\dagger a_{2j}+ a_{3j}^\dagger a_{3j}+2 a_{4j}^\dagger a_{4j} +a_{5j}^\dagger a_{5j} +b_{2j}^\dagger b_{2j} +\tilde b_{2j}^\dagger\tilde b_{2j} \right)\nonumber\\
&&+\left(\frac{1}{4}-u_1 \sin2\alpha\right) \left( b_{1j}^\dagger b_{1j}+b_{4j}^\dagger b_{4j} \right) + \left(\frac{1}{2}-u_1 \sin2\alpha\right) \left( b_{3j}^\dagger b_{3j}+c_j^\dagger c_j   \right)\nonumber\\
&&- \left(\frac{1}{4}+u_1 \sin2\alpha\right) \left(\tilde b_{1j}^\dagger\tilde b_{1j} + \tilde b_{4j}^\dagger\tilde b_{4j}\right) - \left(\frac{1}{2}+u_1 \sin2\alpha\right) \left(\tilde b_{3j}^\dagger\tilde b_{3j} +\tilde c_j^\dagger\tilde c_j  \right)\nonumber\\
&&+\frac{1}{4} \left(2 a_{1j}^\dagger a_{2j}-4 u_2 \cos\alpha a_{4j}^\dagger a_{5j}-4 u_2 \sin\alpha a_{3j}^\dagger a_{4j}+4 u_1 a_{3j}^\dagger a_{5j}+\sqrt{2} b_{1j}^\dagger b_{2j}\right.\nonumber\\
&& -\left.b_{1j}^\dagger b_{4j}+\sqrt{2} b_{2j}^\dagger b_{4j}-\sqrt{2} \tilde b_{1j}^\dagger \tilde b_{2j}+\tilde b_{1j}^\dagger \tilde b_{4j}-\sqrt{2} \tilde b_{2j}^\dagger \tilde b_{4j}+h.c.\right),\nonumber\\
S_{3j}^z &=& -nu_1 \sin2\alpha
-u_2 \cos\alpha \left(P_ja_{2j}+a_{2j}^\dagger P_j\right)
+u_1 \cos2\alpha \left(P_ja_{4j}+a_{4j}^\dagger P_j\right)
-u_2 \sin\alpha \left(P_ja_{5j}+a_{5j}^\dagger P_j\right)\\
&&+u_1 \sin2\alpha \left( a_{1j}^\dagger a_{1j}+a_{2j}^\dagger a_{2j} +a_{3j}^\dagger a_{3j} +2 a_{4j}^\dagger a_{4j}+a_{5j}^\dagger a_{5j} +b_{3j}^\dagger b_{3j} +\tilde b_{3j}^\dagger\tilde b_{3j} \right)\nonumber\\
&&+\left(\frac{1}{4}+u_1 \sin2\alpha\right) \left(b_{1j}^\dagger b_{1j} +b_{4j}^\dagger b_{4j} \right)+\left(\frac{1}{2}+u_1 \sin2\alpha\right) \left(b_{2j}^\dagger b_{2j} +c_j^\dagger c_j   \right)\nonumber\\
&& -\left(\frac{1}{4}-u_1 \sin2\alpha\right) \left(\tilde b_{1j}^\dagger\tilde b_{1j} + \tilde b_{4j}^\dagger\tilde b_{4j}\right) - \left(\frac{1}{2}-u_1 \sin2\alpha\right) \left(\tilde b_{2j}^\dagger\tilde b_{2j} +\tilde c_j^\dagger\tilde c_j  \right)\nonumber\\
&&+ \frac{1}{4} \left(2 a_{1j}^\dagger a_{3j}-4 u_2 \sin\alpha a_{2j}^\dagger a_{4j}+4 u_1 a_{2j}^\dagger a_{5j}+4 u_2 \cos\alpha a_{4j}^\dagger a_{5j}-\sqrt{2} b_{1j}^\dagger b_{3j}\right.\nonumber\\
&& + \left. b_{1j}^\dagger b_{4j}+\sqrt{2} b_{3j}^\dagger b_{4j}+\sqrt{2} \tilde b_{1j}^\dagger \tilde b_{3j}-\tilde b_{1j}^\dagger \tilde b_{4j}-\sqrt{2} \tilde b_{3j}^\dagger \tilde b_{4j}+h.c.\right),\nonumber\\
\label{s4jz}
S_{4j}^z &=& nu_1 \sin2\alpha 
+ u_2 \cos\alpha \left(P_ja_{3j}+a_{3j}^\dagger P_j\right)
-u_1 \cos2\alpha \left(P_ja_{4j}+a_{4j}^\dagger P_j\right)
+u_2 \sin\alpha \left(P_ja_{5j}+a_{5j}^\dagger P_j\right)\\
&&-u_1 \sin2\alpha \left(a_{1j}^\dagger a_{1j} +a_{2j}^\dagger a_{2j} +a_{3j}^\dagger a_{3j} +2 a_{4j}^\dagger a_{4j}+a_{5j}^\dagger a_{5j} +b_{2j}^\dagger b_{2j} +\tilde b_{2j}^\dagger\tilde b_{2j} \right)\nonumber\\
&&+\left(\frac{1}{4}-u_1 \sin2\alpha\right) \left(b_{1j}^\dagger b_{1j} +b_{4j}^\dagger b_{4j} \right)+\left(\frac{1}{2}-u_1 \sin2\alpha\right) \left(b_{3j}^\dagger b_{3j} +c_j^\dagger c_j   \right)\nonumber\\
&& -\left(\frac{1}{4} + u_1 \sin2\alpha\right) \left(\tilde b_{1j}^\dagger\tilde b_{1j} + \tilde b_{4j}^\dagger\tilde b_{4j}\right) - \left(\frac{1}{2}+u_1 \sin2\alpha\right) \left(\tilde b_{3j}^\dagger\tilde b_{3j} +\tilde c_j^\dagger\tilde c_j  \right)\nonumber\\
&&+ \frac{1}{4} \left( 4 u_2 \sin\alpha a_{3j}^\dagger a_{4j}-4 u_1 a_{3j}^\dagger a_{5j}-2 a_{1j}^\dagger a_{2j}-4 u_2 \cos\alpha a_{4j}^\dagger a_{5j}-\sqrt{2} b_{1j}^\dagger b_{2j}\right.\nonumber\\
&& -\left.b_{1j}^\dagger b_{4j}-\sqrt{2} b_{2j}^\dagger b_{4j}+\sqrt{2} \tilde b_{1j}^\dagger \tilde b_{2j}+\tilde b_{1j}^\dagger \tilde b_{4j}+\sqrt{2} \tilde b_{2j}^\dagger \tilde b_{4j}+h.c.\right),\nonumber
\end{eqnarray}

\begin{eqnarray}
\label{s1j+}
S_{1j}^+ &=& \frac{1}{12} \left(2 \sqrt{3} \cos\alpha \cos\beta 
\left( \left(\sqrt{2} b_{3j}^\dagger+2 b_{1j}^\dagger\right)P_j
+ P_j\left(\sqrt{2} \tilde b_{3j}+2 \tilde b_{1j}\right)\right)\right.\\
&& + \sqrt{3} \cos\alpha \sin\beta 
\left( \left(3 \sqrt{2} b_{4j}^\dagger-2 b_{3j}^\dagger+\sqrt{2} b_{1j}^\dagger\right)P_j
+P_j\left(3 \sqrt{2} \tilde b_{4j}-2 \tilde b_{3j}+\sqrt{2} \tilde b_{1j}\right)\right)\nonumber\\
&& + \left. 3 \sin\alpha 
\left(\left(\sqrt{2} b_{4j}^\dagger-2 b_{3j}^\dagger-\sqrt{2} b_{1j}^\dagger\right)P_j
+P_j\left(\sqrt{2} \tilde b_{1j}+2 \tilde b_{3j}-\sqrt{2} \tilde b_{4j}\right)\right)\right)\nonumber\\
&&+
\left(u_3 a_{4j} - \frac{1}{4} \left(2 a_{2j}-\sqrt{6} \cos\beta a_{5j}\right)\right) b_{4j}^\dagger-\frac{1}{\sqrt{2}} \left(a_{1j}+a_{3j}\right) b_{2j}^\dagger + \left(u_4 a_{4j}-u_2 a_{5j}\right) b_{3j}^\dagger\nonumber\\
&&+\left(u_5 a_{4j} - \frac{1}{2} \left(a_{2j}-2 u_6 a_{5j}\right)\right) b_{1j}^\dagger+\left(u_9 a_{4j}^\dagger+\frac{1}{4} \left(2 a_{2j}^\dagger+\sqrt{6} \cos\beta a_{5j}^\dagger\right)\right) \tilde b_{4j}\nonumber\\
&& + \frac{1}{\sqrt2}\left(a_{3j}^\dagger - a_{1j}^\dagger \right) \tilde b_{2j} + \left(u_7 a_{4j}^\dagger-u_2 a_{5j}^\dagger\right) \tilde b_{3j} 
+ \left( \frac{1}{2} a_{2j}^\dagger + u_8 a_{4j}^\dagger + u_6 a_{5j}^\dagger \right) \tilde b_{1j}\nonumber\\
&& + \frac{1}{2} \left(\tilde b_{4j}^\dagger+\sqrt{2} \tilde b_{3j}^\dagger-\tilde b_{1j}^\dagger\right) \tilde c_j + \frac{1}{2} \left(b_{4j}+\sqrt{2} b_{3j}-b_{1j}\right) c_j^\dagger,\nonumber\\
S_{2j}^+ &=& \frac{1}{12} \left(-2 \sqrt{3} \cos\alpha \cos\beta 
\left(\left(\sqrt{2} b_{2j}^\dagger+2 b_{1j}^\dagger\right)P_j
+P_j\left(\sqrt{2} \tilde b_{2j}+2 \tilde b_{1j}\right)\right)\right.\\
&& + \sqrt{3} \cos\alpha \sin\beta 
\left( \left(3 \sqrt{2} b_{4j}^\dagger+2 b_{2j}^\dagger-\sqrt{2} b_{1j}^\dagger \right)P_j
+ P_j\left(3 \sqrt{2} \tilde b_{4j}+2 \tilde b_{2j}-\sqrt{2} \tilde b_{1j}\right)\right)\nonumber\\
&& + \left.3 \sin\alpha 
\left(\left(-\sqrt{2} b_{1j}^\dagger-2 b_{2j}^\dagger-\sqrt{2} b_{4j}^\dagger\right)P_j
+P_j\left(\sqrt{2} \tilde b_{1j}+2 \tilde b_{2j}+\sqrt{2} \tilde b_{4j}\right)\right)\right)\nonumber\\
&& + \left(u_9 a_{4j}+\frac{1}{4} \left(2 a_{3j}+\sqrt{6} \cos\beta a_{5j}\right)\right) b_{4j}^\dagger - \left(u_7 a_{4j}-u_2 a_{5j}\right) b_{2j}^\dagger + \frac{1}{\sqrt{2}}\left(a_{1j}-a_{2j}\right) b_{3j}^\dagger\nonumber\\
&& - \left(\frac{a_{3j}}{2} + u_8 a_{4j} + \frac{1}{2 \sqrt{6}}\left(\cos\beta - 2 \sqrt{2} \sin\beta\right) a_{5j}\right) b_{1j}^\dagger
+ \left(u_3 a_{4j}^\dagger - \frac{1}{4} \left(2 a_{3j}^\dagger - \sqrt{6} \cos\beta a_{5j}^\dagger\right)\right) \tilde b_{4j}\nonumber\\
&& - \left(u_4 a_{4j}^\dagger - u_2 a_{5j}^\dagger\right) \tilde b_{2j} 
+
\frac{1}{\sqrt{2}} \left( a_{1j}^\dagger + a_{2j}^\dagger \right) \tilde b_{3j}+\left(\frac12a_{3j}^\dagger-u_5 a_{4j}^\dagger
- u_6 a_{5j}^\dagger\right) \tilde b_{1j}\nonumber\\
&&+\frac{1}{2} \left(\tilde b_{1j}^\dagger-\sqrt{2} \tilde b_{2j}^\dagger+\tilde b_{4j}^\dagger\right) \tilde c_j+\frac{1}{2} \left(b_{1j}-\sqrt{2} b_{2j}+b_{4j}\right) c_j^\dagger,\nonumber\\
S_{3j}^+ &=& \frac{1}{12} \left(-2 \sqrt{3} \cos\alpha \cos\beta 
\left(\left(\sqrt{2} b_{3j}^\dagger-2 b_{1j}^\dagger\right)P_j
+P_j\left(\sqrt{2} \tilde b_{3j}-2 \tilde b_{1j}\right)\right)\right.\\
&& + 3 \sin\alpha 
\left(\left(\sqrt{2} b_{4j}^\dagger+2 b_{3j}^\dagger-\sqrt{2} b_{1j}^\dagger\right)P_j
-P_j\left(\sqrt{2} \tilde b_{4j}+2 \tilde b_{3j}-\sqrt{2} \tilde b_{1j}\right)\right)\nonumber\\
&& + \left.\sqrt{3} \cos\alpha \sin\beta 
\left( \left(3 \sqrt{2} b_{4j}^\dagger+2 b_{3j}^\dagger+\sqrt{2} b_{1j}^\dagger\right)P_j
+P_j\left(3 \sqrt{2} \tilde b_{4j}+2 \tilde b_{3j}+\sqrt{2} \tilde b_{1j}\right)\right)\right)\nonumber\\
&& + \left(u_3 a_{4j}+\frac{1}{4} \left(2 a_{2j}+\sqrt{6} \cos\beta a_{5j}\right)\right) b_{4j}^\dagger + \frac{1}{\sqrt{2}}\left(a_{1j}-a_{3j}\right) b_{2j}^\dagger - \left(u_4 a_{4j}-u_2 a_{5j}\right) b_{3j}^\dagger\nonumber\\
&& + \left(u_5 a_{4j}+\frac{1}{2} \left(a_{2j} + 2 u_6 a_{5j}\right)\right) b_{1j}^\dagger+\left(u_9 a_{4j}^\dagger - \frac{1}{4} \left(2 a_{2j}^\dagger - \sqrt{6} \cos\beta a_{5j}^\dagger\right)\right) \tilde b_{4j}\nonumber\\
&& + \frac{1}{\sqrt{2}}\left(a_{1j}^\dagger + a_{3j}^\dagger\right) \tilde b_{2j} - \left(u_7 a_{4j}^\dagger - u_2 a_{5j}^\dagger\right) \tilde b_{3j} 
- \left( \frac{1}{2} a_{2j}^\dagger - u_8 a_{4j}^\dagger - u_6 a_{5j}^\dagger \right) \tilde b_{1j}\nonumber\\
&&+\frac{1}{2} \left(\tilde b_{4j}^\dagger-\sqrt{2} \tilde b_{3j}^\dagger-\tilde b_{1j}^\dagger\right) \tilde c_j+\frac{1}{2} \left(b_{4j}-\sqrt{2} b_{3j}-b_{1j}\right) c_j^\dagger,\nonumber\\
\label{s4j+}
S_{4j}^+ &=& \frac{1}{12} \left(2 \sqrt{3} \cos\alpha \cos\beta 
\left(\left(\sqrt{2} b_{2j}^\dagger-2 b_{1j}^\dagger\right)P_j
+P_j\left(\sqrt{2} \tilde b_{2j}-2 \tilde b_{1j}\right)\right)\right.\\
&& - 3 \sin\alpha 
\left( \left(\sqrt{2} b_{1j}^\dagger - 2 b_{2j}^\dagger + \sqrt{2} b_{4j}^\dagger\right)P_j 
-P_j\left(\sqrt{2} \tilde b_{1j} - 2 \tilde b_{2j} + \sqrt{2} \tilde b_{4j}\right)\right)\nonumber\\
&& +\left.\sqrt{3} \cos\alpha \sin\beta 
\left( \left(3 \sqrt{2} b_{4j}^\dagger - 2 b_{2j}^\dagger - \sqrt{2} b_{1j}^\dagger\right)P_j 
+ P_j\left(3 \sqrt{2} \tilde b_{4j} - 2 \tilde b_{2j} - \sqrt{2} \tilde b_{1j}\right)\right)\right)\nonumber\\
&&+\left(u_9 a_{4j} - \frac{1}{4} \left( 2 a_{3j} - \sqrt{6} \cos\beta a_{5j}\right)\right) b_{4j}^\dagger+\left(u_7 a_{4j}-u_2 a_{5j}\right) b_{2j}^\dagger
-\frac{1}{\sqrt{2}}\left(a_{1j}+a_{2j}\right) b_{3j}^\dagger\nonumber\\
&&+\left(\frac{a_{3j}}{2}-u_8 a_{4j} 
-u_6 a_{5j}\right) b_{1j}^\dagger
+\left(u_3 a_{4j}^\dagger+\frac{1}{4} \left(2 a_{3j}^\dagger+\sqrt{6} \cos\beta a_{5j}^\dagger\right)\right) \tilde b_{4j}\nonumber\\
&& +\left(u_4 a_{4j}^\dagger-u_2 a_{5j}^\dagger\right) \tilde b_{2j}
- \frac{1}{\sqrt{2}}\left( a_{1j}^\dagger - a_{2j}^\dagger \right) \tilde b_{3j}
- \left( \frac12 a_{3j}^\dagger + u_5 a_{4j}^\dagger + u_6 a_{5j}^\dagger
\right) \tilde b_{1j}\nonumber\\
&&+\frac{1}{2} \left(\tilde b_{1j}^\dagger+\sqrt{2} \tilde b_{2j}^\dagger+\tilde b_{4j}^\dagger\right) \tilde c_j+\frac{1}{2} \left(b_{1j}+\sqrt{2} b_{2j}+b_{4j}\right) c_j^\dagger,\nonumber
\end{eqnarray}
\begin{eqnarray}
&&\frac1n\left({\bf S}_{1j}{\bf S}_{2j} + {\bf S}_{2j}{\bf S}_{3j} + {\bf S}_{3j}{\bf S}_{4j} + {\bf S}_{4j}{\bf S}_{1j}\right) = 
-\frac{n}{4} \left(3-\cos2\alpha+6 \cos^2\alpha \cos2\beta\right)\\
&&+\left(\frac{1}{4} (1-3 \cos2\beta) \sin2\alpha \left(P_ja_{4j}+a_{4j}^\dagger P_j\right)+\frac{3}{2} \cos\alpha \sin2\beta \left(P_ja_{5j}+a_{5j}^\dagger P_j\right)\right)\nonumber\\
&&+ \frac{1}{4} \left(3-\cos2\alpha+6 \cos^2\alpha \cos2\beta\right) \left(a_{1j}^\dagger a_{1j} +  a_{2j}^\dagger a_{2j} +a_{3j}^\dagger a_{3j} \right)\nonumber\\
&& -\frac{1}{2} \cos2\alpha (1-3 \cos2\beta) a_{4j}^\dagger a_{4j}+\frac{1}{4} \left(3 (3+\cos2\alpha) \cos2\beta+2 \sin^2\alpha\right) a_{5j}^\dagger a_{5j}
\nonumber\\
&&+3 \cos\beta \sin\alpha \sin\beta \left(a_{4j}^\dagger a_{5j}+a_{4j} a_{5j}^\dagger\right)+\frac{1}{4} \left(7-\cos2\alpha+6 \cos^2\alpha \cos2\beta\right) \left(b_{4j}^\dagger b_{4j} + \tilde b_{4j}^\dagger\tilde b_{4j}+\tilde c_j^\dagger\tilde c_j  +c_j^\dagger c_j   \right)
\nonumber\\
&&+\frac{1}{4} \left(3-\cos2\alpha+6 \cos^2\alpha \cos2\beta\right) \left(b_{2j}^\dagger b_{2j} + b_{3j}^\dagger b_{3j}+\tilde b_{2j}^\dagger\tilde b_{2j} +\tilde b_{3j}^\dagger\tilde b_{3j} \right)\nonumber\\
&& -\frac{1}{2} \cos^2\alpha (1-3 \cos2\beta) \left(b_{1j}^\dagger b_{1j} +\tilde b_{1j}^\dagger\tilde b_{1j} \right), \nonumber
\end{eqnarray}
\begin{equation}
\frac1n\left({\bf S}_{1j}{\bf S}_{3j} + {\bf S}_{2j}{\bf S}_{4j}\right) = 
\frac{n}{2} 
- \left(
2 a_{1j}^\dagger a_{1j} + a_{2j}^\dagger a_{2j} + a_{3j}^\dagger a_{3j} + b_{2j}^\dagger b_{2j} + b_{3j}^\dagger b_{3j} + \tilde b_{2j}^\dagger \tilde b_{2j} + \tilde b_{3j}^\dagger \tilde b_{3j}
\right),
\end{equation}
where $P_j$ is given by Eq.~\eqref{proj4} and
\begin{eqnarray}
u_1 &=& \frac{1}{2\sqrt3} (\sin\beta-\sqrt{2} \cos\beta),\\
u_2 &=& \frac{1}{2\sqrt3} (\cos\beta+\sqrt{2} \sin\beta),\\
u_3 &=& \frac{1}{2\sqrt2} (\sqrt{3} \sin\alpha \sin\beta - \cos\alpha),\\
u_4 &=& \frac{1}{2\sqrt3} (\sqrt{3} \cos\alpha+\sqrt{2} \cos\beta \sin\alpha-\sin\alpha \sin\beta),\\
u_5 &=& \frac{1}{2\sqrt6} (\sqrt{3} \cos\alpha+2 \sqrt{2} \cos\beta \sin\alpha+\sin\alpha \sin\beta),\\
u_6 &=& \frac{1}{2\sqrt6} (\cos\beta-2 \sqrt{2} \sin\beta),\\
u_7 &=& \frac{1}{2\sqrt3} (\sqrt{2} \cos\beta \sin\alpha-\sqrt{3} \cos\alpha-\sin\alpha \sin\beta),\\
u_8 &=& \frac{1}{2\sqrt6} (2 \sqrt{2} \cos\beta \sin\alpha-\sqrt{3} \cos\alpha+\sin\alpha \sin\beta),\\
u_9 &=& \frac{1}{2\sqrt2} (\cos\alpha+\sqrt{3} \sin\alpha \sin\beta).
\end{eqnarray}
Representations for $S^-$ operators are obtained from Eqs.~\eqref{s1j+}--\eqref{s4j+} by Hermitian conjugation.

Raman operator discussed in Sec.~\ref{ramansec} contains the following combination:
\begin{eqnarray}
\label{b1gbond}
&&{\bf S}_{1j}{\bf S}_{2j} - {\bf S}_{2j}{\bf S}_{3j} + {\bf S}_{3j}{\bf S}_{4j} - {\bf S}_{4j}{\bf S}_{1j} = 
\sqrt{3} \cos \alpha  \cos \beta  \left(P_j a_{1j}+a_{1j}^\dagger P_j \right) \\
&&{} - a_{2j}^\dagger a_{3j} - a_{3j}^\dagger a_{2j} +
\sqrt{3} \cos \beta  \sin \alpha  \left(a_{1j}^\dagger a_{4j} + a_{4j}^\dagger a_{1j} \right)
- \sqrt{3} \sin \beta \left(a_{1j}^\dagger  a_{5j} + a_{5j}^\dagger a_{1j} \right) \nonumber\\
&&{} - b_{2j}^\dagger  b_{3j} - b_{3j}^\dagger b_{2j} 
- \tilde b_{2j}^\dagger  \tilde b_{3j} - \tilde b_{3j}^\dagger \tilde b_{2j}. \nonumber
\end{eqnarray}

\bibliography{BondRep}

\end{document}